%% file: main.tex
\documentclass{usiinftr}

\usepackage{babel}
\usepackage{blindtext}
\usepackage{latexsym}

\usepackage{booktabs} 

\usepackage{xcolor}
\definecolor{pennblue}{cmyk}{1,.65,0,.3}
\definecolor{pennred}{cmyk}{0,1,.65,.34}

\usepackage{subcaption}

\usepackage{graphicx}
\graphicspath{{./figures/}{../figures/}}

\usepackage{amsmath}
\usepackage{alltt}

\usepackage{url}

\usepackage{algorithm}
\usepackage[noend]{distribalgo}
\newcommand{\mv}[1]{\emph{#1}}

\begin{document}
\title{Partitioned Paxos via the Network Data Plane}

\author{Huynh Tu Dang}{1}
\author{Pietro Bressana}{1}
\author{Han Wang}{2}
\author{Ki Suh Lee}{3}
\author{Noa Zilberman}{4}
\author{Hakim Weatherspoon}{5}
\author{Marco Canini}{6}
\author{Fernando Pedone}{1}
\author{Robert Soul\'{e}}{1}

\affiliation{1}{\USIINF}
\affiliation{2}{Barefoot Networks, United States}
\affiliation{3}{Waltz Networks, United States}
\affiliation{4}{Computer Laboratory, University of Cambridge, United Kingdom}
\affiliation{5}{Department of Computer Science, Cornell University, United States}
\affiliation{6}{Computer Science, King Abdullah University of Science and Technology, Saudi Arabia}

\TRnumber{2019-01}

\date{}
\thispagestyle{empty}

\maketitle

\begin{abstract}
\input{abstract}

\end{abstract}

\section{Introduction}

Consensus protocols are used to solve a fundamental problem in
distributed systems: getting a group of participants to reliably
agree on some value.  They are the foundation for building
fault-tolerant, distributed applications and services (e.g.,
OpenReplica~\cite{openreplica}, Ceph~\cite{ceph}, Google's
Chubby~\cite{burrows06}). Moreover, many classic distributed systems
problems can be reduced to consensus, including atomic
broadcast~\cite{reed08} and atomic commit~\cite{gray06}.
Unfortunately, consensus protocols are also widely acknowledged
as a performance bottleneck, causing many systems to eschew
strong consistency~\cite{vogels09}. Twenty years ago, researchers
cautioned against using consensus in-band for systems with high
demand~\cite{friedman96}. Still, despite two decades of research on
optimizations~\cite{moraru13,Benz:2014,Lamport04,PS99,lamport06,ports15,marandi10,reed08},
consensus performance remains a problem in real-world systems~\cite{kulkarni15}.

Recently, several projects~\cite{ports15,istvan16,li16,jin18} have
explored a promising new approach to achieving high-performance
consensus. These systems leverage the emerging trend of programmable
networks hardware~\cite{jose15,bosshart13,xpliant} to optimize
consensus, achieving eye-popping performance results.  For example, NoPaxos~\cite{li16} use a Cavium network processor to enforce
ordered message delivery, and reaches a throughput of around 250K messages per second with a no-op, closed-loop client---a 370\%
increase over a standard Paxos baseline.
An implementation of Chain Replication~\cite{vanrenesse04} on a Tofino ASIC~\cite{jin18} is able to process commands at a throughput of 4 billion messages per second, which is several orders of magnitude greater than a software-only alternative.

But, while network-accelerated consensus shows great promise, current
systems suffer from an important limitation: they do not address how
the replicated application can cope with the increased rate of
consensus messages.  For example, the aforementioned
NoPaxos~\cite{li16}, when used to replicate a transactional key-value
store, can only achieve a throughput of 13K transactions per second.
Improving the performance in the network is not sufficient. By solving one bottleneck,
a new one is revealed at the host, in the replicated application.

Prior work such as Consensus in a Box~\cite{istvan16} and
NetChain~\cite{jin18} sidestep this issue to some extent, by
implementing the replicated application itself in network hardware
(i.e., both systems implement a key-value store in their target
hardware devices).  This approach severely limits the applicability of
in-network consensus, as the network really provides a specialized
replicated service, rather than a general-purpose, high-performance
consensus that can be used by any off-the-shelf application.

The usefulness of in-network computing becomes questionable if the
application can not take advantage of the performance it delivers,
especially as network acceleration comes at a cost, in terms of
money, power consumption, or design time.  Indeed, the main
problem that motivates this paper is that \emph{even if consensus
protocols can execute at 6.5Tbps, the exercise is purely academic if
replicated applications cannot cope with that rate of consensus
messages.}

To address this problem, this paper proposes Partitioned Paxos, a
novel approach to network-accelerated consensus. Partitioned Paxos is
based on the observation that there are two aspects to consensus,
\emph{execution} and \emph{agreement}.
Execution governs how replicas execute transitions in a state machine,
while agreement ensures that replicas execute transitions in the same
order. Consensus in a Box~\cite{istvan16} and NetChain~\cite{jin18}
perform both execution and agreement inside the network.  In contrast,
the key insight behind Partitioned Paxos is to isolate and separately
optimize these two concerns.  This allows any application to
take advantage of optimized consensus.

Partitioned Paxos uses programmable network hardware to accelerate
agreement, following Lamport's Paxos algorithm~\cite{lamport98}.
Thus, it accelerates consensus protocols without strengthening
assumptions about the behavior of the network. Then, to leverage the
increased rate of consensus and optimize execution, Partitioned Paxos
shards the application state, and runs parallel Paxos deployments for
each shard. By sharding the state of the application, we multiply the
performance of the application by the number of partitions/shards.
Overall, our solution indicates that the only way to significantly
improve application performance is through close hardware/software
co-design, and an across-stack optimization of all the components, from the network, through network stack (kernel bypass), file system
and storage.

Partitioned Paxos provides significant performance improvements
compared with traditional implementations.  In a data center network,
Partitioned Paxos reduces the latency by $\times 3$. In terms of
agreement throughput, our implementation on Barefoot Network's Tofino
ASIC chip~\cite{bosshart13} can process over 2.5 billion consensus
messages per second, a four-order-of-magnitude improvement. In terms
of execution, we have used Partitioned Paxos to accelerate an
unmodified instance of RocksDB. When using 4 separate partitions, the
replicated application can process close to 600K messages per second, a
$\times 11$ improvement over baseline implementations.

In short, this paper makes the following contributions:

\begin{itemize}
\item It describes a novel approach to network-accelerated consensus
that separates agreement from execution.
\item It describes a re-interpretation of the Paxos protocol that maps the consensus protocol logic into stateful forwarding
decisions.
\item It discusses a technique for partitioning and parallelizing replica
state and execution. 
 \item It presents an open-source implementation of Paxos with at least $\times 3$ latency improvement and
 $\times 11$ throughput improvement for unmodified applications.
\end{itemize}

 This paper first provides
 evidence of performance obstacles in Paxos
 (\S\ref{sec:background}). It then presents the Partitioned Paxos
 design, focusing on how the system accelerates agreement
 (\S\ref{sec:agreement}) and execution (\S\ref{sec:execution}).  Next,
 it presents an evaluation on programmable ASICs
 (\S\ref{sec:evaluation}). Related work is discussed in
 (\S\ref{sec:related}). Finally, it concludes in
 (\S\ref{sec:conclusion}).

\section{Paxos Bottlenecks and Solution Overview}
\label{sec:background}

Before presenting the design of Partitioned Paxos, we briefly
review background on Paxos and the obstacles
for achieving high-performance consensus.

\subsection{Paxos Overview}
Paxos~\cite{lamport98} is a consensus protocol that makes very few
assumptions about the network behavior (e.g., point-to-point packet
delivery and the election of a non-faulty leader), making it widely
applicable to a number of deployment scenarios. It has been
proven safe under asynchronous assumptions, live under weak
synchronous assumptions, and resilience-optimum~\cite{Lam06b}.


Paxos distinguishes the following roles that a process can play: {\em
proposers}, {\em acceptors} and {\em replicas}. The proposers submit commands that
need to be ordered by Paxos before they are learned and executed by
the replicated state machines. The acceptors are the processes that actually agree on a value.


An instance of Paxos proceeds in two phases. During Phase 1, a
proposer that wants to submit a value selects a unique round number
and sends a prepare request to at least a quorum of acceptors.
Upon receiving a prepare request with a round number bigger
than any previously received round number, the acceptor responds to
the proposer promising that it will reject any future requests
with smaller round numbers.  If the acceptor already accepted a
request for the current instance, it will return the
accepted value to the proposer, together with the round number
received when the request was accepted. When the proposer receives
answers from a quorum of acceptors, the second phase begins.

In Phase 2, the proposer selects a value
according to the following rule.  If no value is returned in the
responses, the proposer can select a new value for
the instance; however, if any of the acceptors returned a value in the
first phase, the proposer must select the value with the highest round
number among the responses. The proposer then sends an accept request with the round number used in the first phase and the value selected to at least a
quorum of acceptors.  When receiving such a request, the acceptors
acknowledge it by sending the accepted value to the replicas, unless the
acceptors have already acknowledged another request with a higher
round number.  When a quorum of acceptors accepts a value, consensus is reached.

Once consensus is reached, the accepted value is delievered
to the application. Usually, the value is a state-machine transition
that will modify the application state (e.g., write a value).
As will be discussed in Section~\ref{sec:execution},
one of the Partitioned Paxos optimizations depends on
sharding or partitioning this state.

If multiple proposers simultaneously execute the procedure above for
the same instance, then no proposer may be able to execute the two
phases of the protocol and reach consensus.  To avoid scenarios in
which proposers compete indefinitely, a
\emph{leader} process can be elected.  Proposers submit values to the leader, which executes the first and second
phases of the protocol.  If the leader fails, another process takes
over its role.  Paxos ensures consistency despite concurrent leaders
and progress in the presence of a single leader.

In practice, replicated services run multiple executions of the Paxos
 protocol to achieve consensus on a sequence of
 values~\cite{chandra07} (i.e., multi-Paxos). An execution of Paxos is
 called an instance. In this paper, we implicitly describe a
 multi-Paxos protocol.

\subsection{Performance Obstacles}

Given the central role that Paxos plays in fault-tolerant, distributed
systems, improving the performance of the protocol has been an intense
area of
study~\cite{moraru13,Benz:2014,Lamport04,PS99,lamport06,ports15,marandi10,reed08}.
There are obstacles related to the protocol itself, in
terms of latency and throughput, and related to the application.

\paragraph{Protocol Latency.}
The performance of Paxos is typically measured in ``communication
steps'', where a communication step corresponds to a server-to-server
communication in an abstract distributed system.  Lamport proved that
it takes at least 3 steps to order messages in a distributed setting~\cite{Lam06b}. This means that there is not much hope for
significant performance improvements, unless one revisits the model
(e.g., \cite{charron-bost04}) or assumptions
(e.g., \emph{spontaneous message ordering}~\cite{lamport06, PS02b,
PSUC02}).

\begin{figure}[t]
 \centering
  \includegraphics[width=0.5\columnwidth]{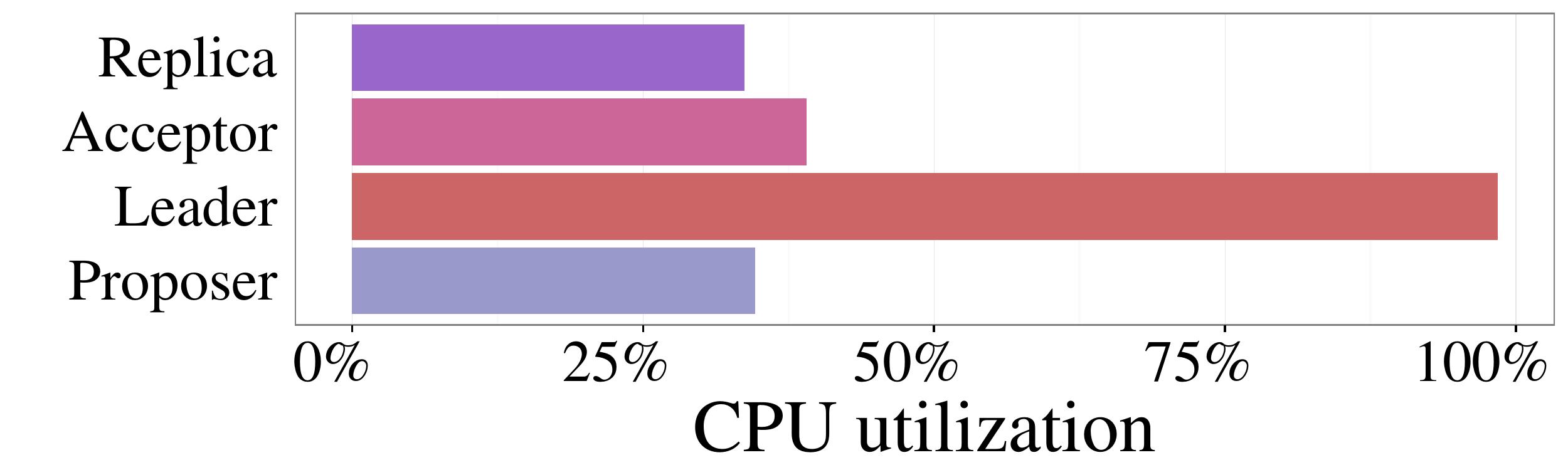}
\caption{The leader process becomes a bottleneck at a throughput of \textasciitilde65K values/second.}
  \label{fig:bottleneck1}
   \vspace{-1em}
\end{figure}

\paragraph{Protocol Throughput.}
Beyond latency due to communication steps, throughput is also
a challenge for Paxos.  Under load, a Paxos leader becomes a
bottleneck~\cite{li16}, since the role interposes on all messages that proposers submit.
To demonstrate, we performed a basic experiment in which we measured the CPU utilization for each
of the Paxos roles when transmitting messages at peak throughput.  As a representative implementation of Paxos, we used the open-source \texttt{libpaxos} library~\cite{libpaxos}. We chose \texttt{libpaxos}
because it is a faithful implementation of Paxos that distinguishes
all the Paxos roles. And, it has been extensively tested and is often
used as a reference implementation (e.g., \cite{istvan16,marandi14,Poke:2015,sciascia13}).
In the experiment, a client application sent 64-byte messages at increasing
rates until we reached the saturation point at a peak throughput rate of $\sim$65K values/sec. As shown in
Figure~\ref{fig:bottleneck1}, the leader is the first process to become CPU bound.

\begin{figure}[t]
  \centering
  \includegraphics[width=0.4\columnwidth]{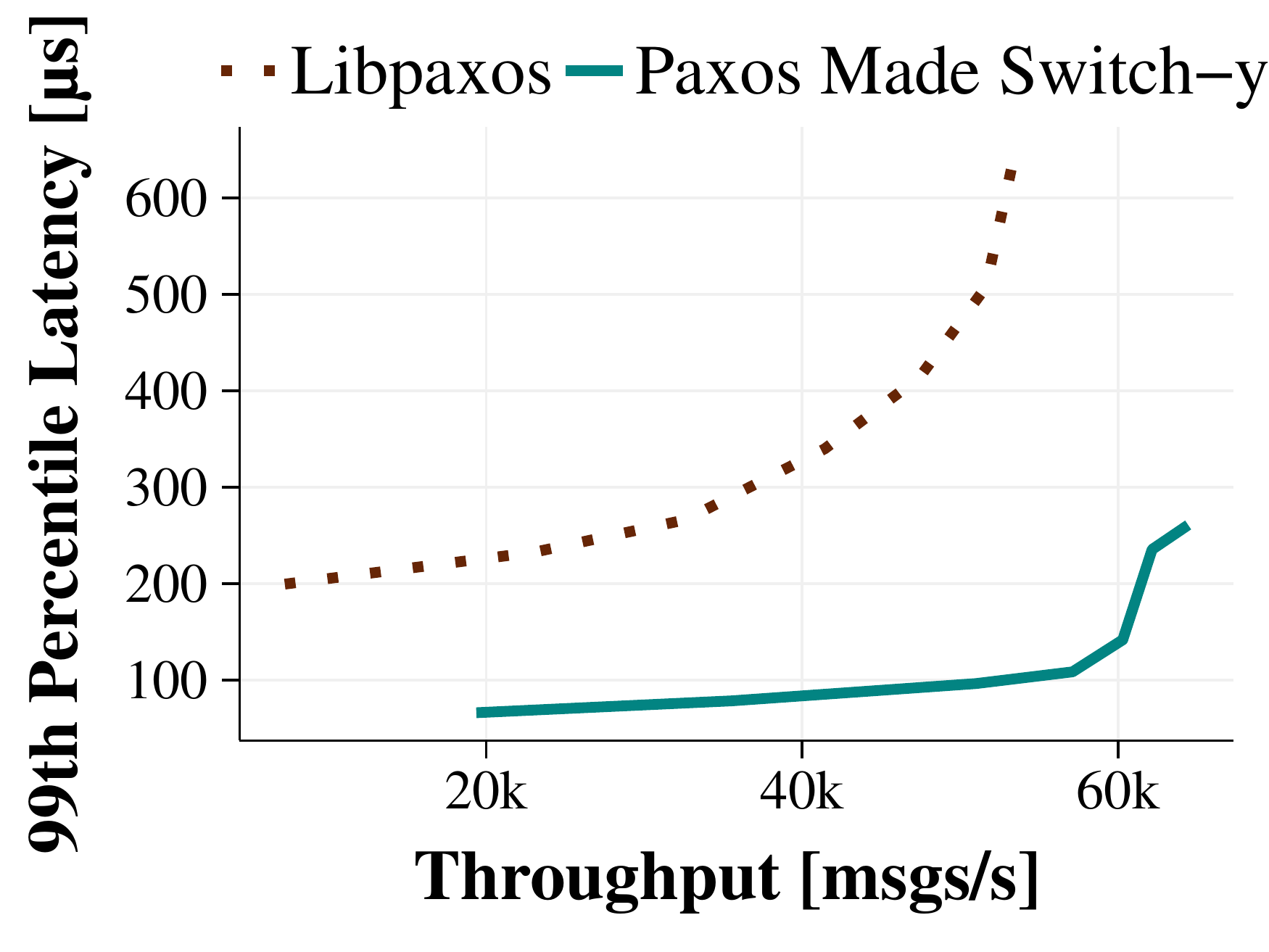}
  \caption{Throughput vs. latency (w/ RocksDB).}
 \vspace{-1em}
\label{fig:tput-vs-latency-rocksdb}
\end{figure}

\paragraph{Application Bottleneck.}
To ensure correctness, replicas must execute commands in a deterministic order.  This ordering becomes a
bottleneck, even if the rest of the Paxos protocol can be accelerated.
This problem has been mentioned, or alluded to, in prior work~\cite{li16,istvan16,jin18}.
To further illustrate the point, we implemented Lamport's Paxos in P4, following the design in
Paxos made Switch-y~\cite{dang16}. We ran the program on a switch with Barefoot Network's Tofino ASIC
chip~\cite{bosshart13}, and used the protocol to replicate an unmodified instance of RocksDB.
Figure~\ref{fig:tput-vs-latency-rocksdb} plots the latency vs. throughput for
deployments using \texttt{libpaxos} and the P4 Paxos implementation. While a network implementation of Paxos certainly improves the performance, increasing the throughput from $\sim$54K messages per second to $\sim$65K messages per second, the performance gains are a far cry from what seems possible.
The limiting factor becomes the replica.





\subsection{Solution Approach}

Paxos is a protocol for implementing state-machine replication~\cite{schneider90}.  Each replica is a state-machine that makes a transition based on a given input and current state. Paxos requires $f+1$ replicas to tolerate $f$ failures.  However, Paxos requires an additional $2f+1$ processes (because of the majority
voting among acceptors) to ensure that the replicas execute transitions in the same order, despite potential message loss. These two aspects of the protocol are referred to as \emph{execution} and \emph{agreement}, respectively.

Existing approaches to network-accelerated consensus~\cite{istvan16,jin18} optimize both execution and agreement
by offloading to hardware. In contrast, Partitioned Paxos uses two separate techniques for optimizing the different aspects of Paxos. First, it uses the network forwarding plane to accelerate the agreement components of
Paxos. Then, it uses state partitioning and parallelization to accelerate the performance of the replicas. As a result, replicated applications can leverage the performance provided by in-network acceleration and multiple threads to implement strongly consistent services that make only weak assumptions about the network. In our experiments, we have used Partitioned Paxos to replicate an unmodified instance of RocksDB, a production quality key-value store used at Facebook, Yahoo!, and LinkedIn.

\section{Accelerating Agreement}
\label{sec:agreement}

Partitioned Paxos accelerates consensus by moving some of the logic---the agreement aspect---into the network forwarding plane. This addresses two of the major obstacles for achieving high-performance. First, it avoids network I/O bottlenecks in software implementations. Second, it reduces end-to-end latency by executing consensus logic as messages pass through the network.

However, accelerating consensus is not as straightforward as simply ``implementing Paxos on faster hardware''. There are several challenges that arise when considering how to execute consensus
protocols in the network forwarding plane: (i) What is the expected deployment? (ii) How do you map the protocol into the match-action abstractions exposed by network hardware? (iii) How is the failure model of Paxos impacted? and (iv) How do the limited resources in network hardware impact the protocol? Below, we discuss these issues in detail.

\subsection{Deployment}

\begin{figure}
\centering
 \includegraphics[width=0.5\columnwidth]{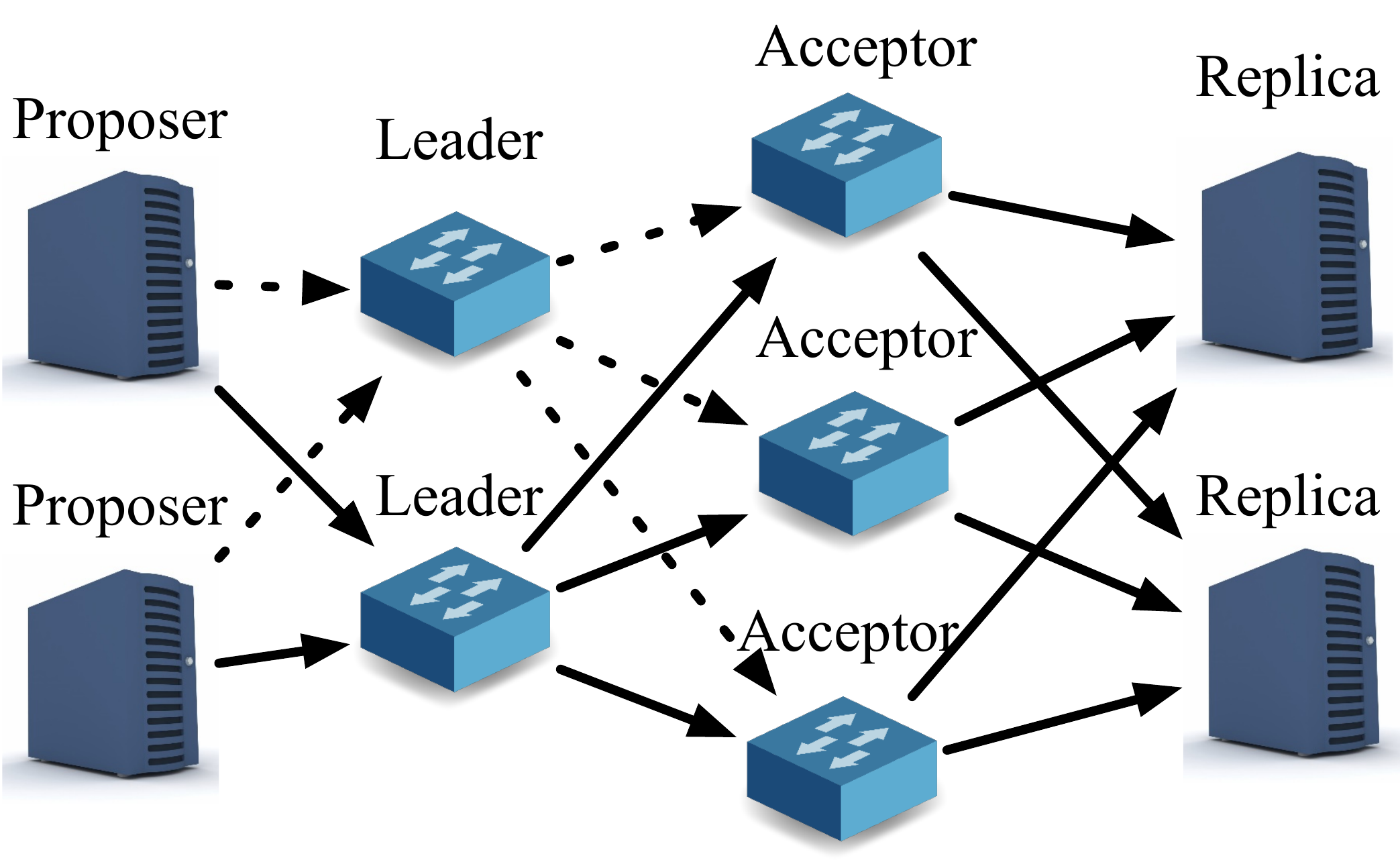}
\caption{Example deployment for Partitioned Paxos.}
\label{fig:deployment}
 \vspace{-1em}
\end{figure}

Figure~\ref{fig:deployment} illustrates a minimal deployment
for Partitioned Paxos. With Partitioned Paxos, network switches
execute the logic for the leader and acceptor roles in Paxos.
The hosts serve as proposers and replicated applications.

For availability, Paxos intrinsically assumes that if a node fails, the
other nodes can still communicate with each other. By moving this logic
into network devices, Partitioned Paxos necessarily mandates that
there are redundant communication paths between devices. In
Figure~\ref{fig:deployment}, a redundant path between proposers
and a backup leader is illustrated with dashed lines.
In a more realistic, data center deployment, this redundancy is
already present between top-of-rack (ToR), aggregate, and spine switches.

\begin{figure}
\centering
 \includegraphics[width=0.7\columnwidth]{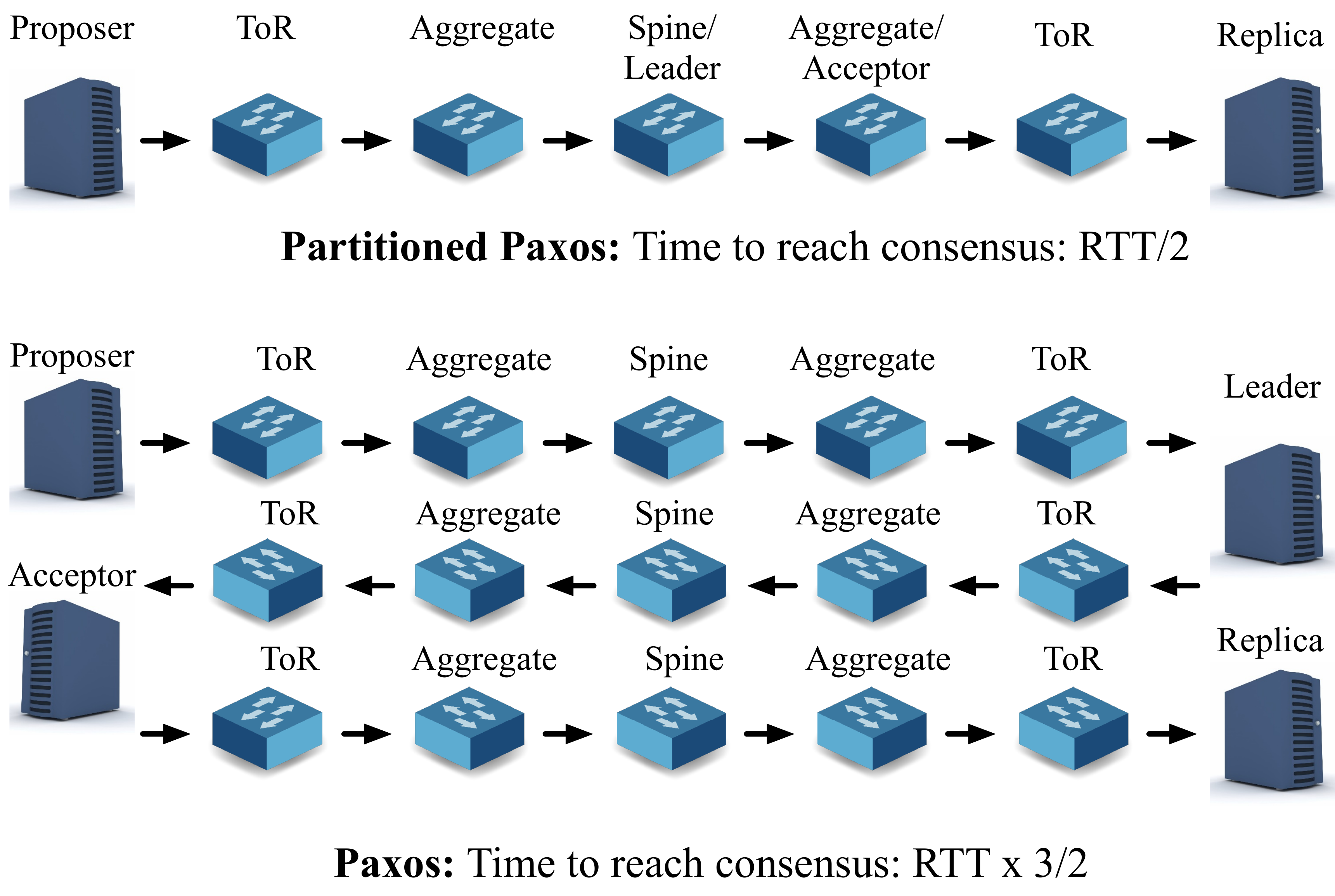}
\caption{Contrasting propagation time for Partitioned Paxos with server-based deployment.}
\label{fig:p4xos}
 \vspace{-1em}
\end{figure}

Figure~\ref{fig:p4xos} illustrates the difference in number of hops
needed by Partitioned Paxos and traditional deployments.
 While in a standard Paxos implementation, every communication step
requires traversing the network (e.g., Fat-tree), in Partitioned
Paxos, each network device fills a role in achieving consensus. Note
that the number of uplinks from a ToR does not need to reach a quorum
of aggregators, as the aggregator may also reside at the fourth hop.
Partitioned Paxos saves two traversals of the network compared to
Paxos, meaning $\times 3$ latency improvement.

In absolute numbers, this reduction
is significant.  Our experiments show that the Paxos logic execution
time takes around 2.5us, without I/O. Using kernel-bypass~\cite{dpdk},
a packet can be sent out of host in $\sim$5us
(median)~\cite{zilberman2017time}. One way delay in the data center is
$\sim$100us (median)~\cite{popescu2017ptpmesh}, more than $\times 10$ the host!
Implementing Paxos in switch ASICS as ``bumps-in-the-wire'' processing
allows consensus to be reached in sub-round-trip time (RTT).

It is worth mentioning that this deployment does not require
additional hardware to be deployed in the network, such as Middleboxes
or FPGAs. Partitioned Paxos leverages hardware resources that are
already available.




\subsection{Paxos as Match-Action}

Conceptually, implementing Paxos in network hardware involves mapping
the logic into a sequence of match-action units. Perhaps surprisingly,
Paxos logic maps well into this abstraction.  In the pseudocode below,
we describe the protocol as responses to different input messages. In
other words, we re-interpret the Paxos algorithm as a set of stateful
forwarding decisions.  This presentation of the algorithm presents a
different way of understanding the notoriously complex and subtle
protocol~\cite{lamport01,mazieres07,vanrenesse15,chandra07}.

Prior work on Paxos made Switch-y~\cite{dang16} presented P4 code for the Phase
2 logic of Paxos leaders and acceptors. Partitioned
Paxos extends their approach in two key ways. First, Partitioned Paxos
implements both Phase 1 and Phase 2 of the protocol.  Second, Partitioned Paxos targets an ASIC
deployment, which imposes new constraints on the implementation,
including how to trim the acceptor log.

Beyond the presentation, Partitioned Paxos differs from a standard
Paxos deployment in that each command that is submitted must include
the partition identifier. A partition identifier corresponds to a
shard of the application state. Many distributed systems,
such as key-value stores or databases, are naturally partitioned,
e.g., by key-space. We refer to these partitions as application state shards.
Each shard is processed by a parallel
deployment of Paxos. Thus, the partition identifier is threaded
through all of the pseudocode.

\paragraph{Notation.}
Our pseudocode roughly correspond to P4 statements.
The \texttt{Initialize} blocks identify state
 stored in registers. \texttt{id[N]} indicates
 a register named \texttt{id} with \texttt{N} cells.
The notation ``$:= \{0\}$''
 indicates that every cell element in the register should be
 initialized to 0. The \texttt{match} blocks correspond to table
 matches on a packet header, and the \texttt{case} blocks correspond
 to P4 actions. We distinguish updates to the local state (``$:=$''), from
 writes to a packet header (``$\leftarrow$''). We
 also distinguish between unicast (\texttt{forward}) and
multicast (\texttt{multicast}).

\input{algorithms/algorithm-leader}

\paragraph{Partitioned Paxos Packets.}
The Partitioned Paxos packet header includes six fields.  The six
fields are as follows: $(i)$ \texttt{msgtype} distinguishes the
various Paxos messages
(e.g., \texttt{REQUEST}, \texttt{PHASE1A}, \texttt{PHASE2A}, etc.)
$(ii)$ \texttt{inst} is the consensus instance number; $(iii)$
\texttt{rnd} is either the round number computed by the proposer or the round
number for which the acceptor has cast a vote; \texttt{vrnd} is the
round number in which an acceptor has cast a vote;
$(iv)$ \texttt{swid} identifies the sender of the message;
The \texttt{pid} is used to identify the partition; and
$(v)$ \texttt{value} contains the request from the proposer or the
value for which an acceptor has cast a vote.  Our prototype requires
that the entire Paxos header, including the value, be less than the
maximum transmission unit of 1500 bytes.
Note that the proposer is responsible for populating the header.
The client application simply provides the \texttt{value}.


\paragraph{Proposer.}
A Partitioned Paxos proposer mediates client requests, and
encapsulates the request in a Paxos header. It is implemented as a
user-space library that exposes a small API to client applications.

The Partitioned Paxos proposer library is a drop-in replacement for
existing software libraries. The API consists of a
single \texttt{submit} function. The \texttt{submit} function is
called when the application uses Paxos to send a value. The
application simply passes a character buffer containing the value, and
the buffer size.

When a Partitioned Paxos proposer submits a command, it must include
the partition identifier.  The proposer library adds the partition id
to each Paxos command.  The id is not exposed to the client
application.

We note that an optimal sharding of application state is dependent on the workload.
Our prototype uses an even distribution of the key-space.
Determining an optimal sharding of application state is an orthogonal
problem, and an interesting direction for future work.

\input{algorithms/algorithm-acceptor}

\paragraph{Leader.}
A leader brokers requests on behalf of proposers.  The leader ensures
that only one process submits a message to the protocol for a
particular instance (thus ensuring that the protocol terminates), and
imposes an ordering of messages.  When there is a single leader, a
monotonically increasing sequence number can be used to order the
messages.  This sequence number is written to the \texttt{inst} field
of the header.

Algorithm~\ref{fig:leader} shows the pseudocode for the primary leader
implementation.  The leader receives \texttt{REQUEST} messages from
the proposer. \texttt{REQUEST} messages only contain a value. The
leader must perform the following: for a given partition, write the current
instance number and an initial round number into the message header;
increment the instance number for that partition for the next invocation; store the value
of the new instance number; and broadcast the packet to acceptors.

Partitioned Paxos uses a well-known Paxos optimization~\cite{gray06}, where each instance is reserved for the primary leader at initialization (i.e., round number zero).
Thus, the primary leader does not need to execute Phase 1 before
submitting a value (in a \texttt{REQUEST} message) to the acceptors.
Since this optimization only works for one leader, the backup leader
must reserve an instance before submitting a value to the acceptors.
To reserve an instance, the backup leader must send a unique round
number in a \texttt{PHASE1A} message to the acceptors.  For brevity,
we omit the backup leader algorithm since it essentially follows the
Paxos protocol.

\paragraph{Acceptor.}
Acceptors are responsible for choosing a single value for a particular
instance. For each instance of consensus, each individual acceptor must ``vote''
for a value. Acceptors must maintain and access the history of proposals for
which they have voted. This history ensures that acceptors never vote for
different values for a particular instance, and allows the protocol to tolerate
 lost or duplicate messages. This history is referred to as the acceptor log.
The acceptor log must be periodically trimmed, which we describe in
Section~\ref{sec:resource-constraints}.

\begin{figure}
\centering
 \includegraphics[width=0.5\columnwidth]{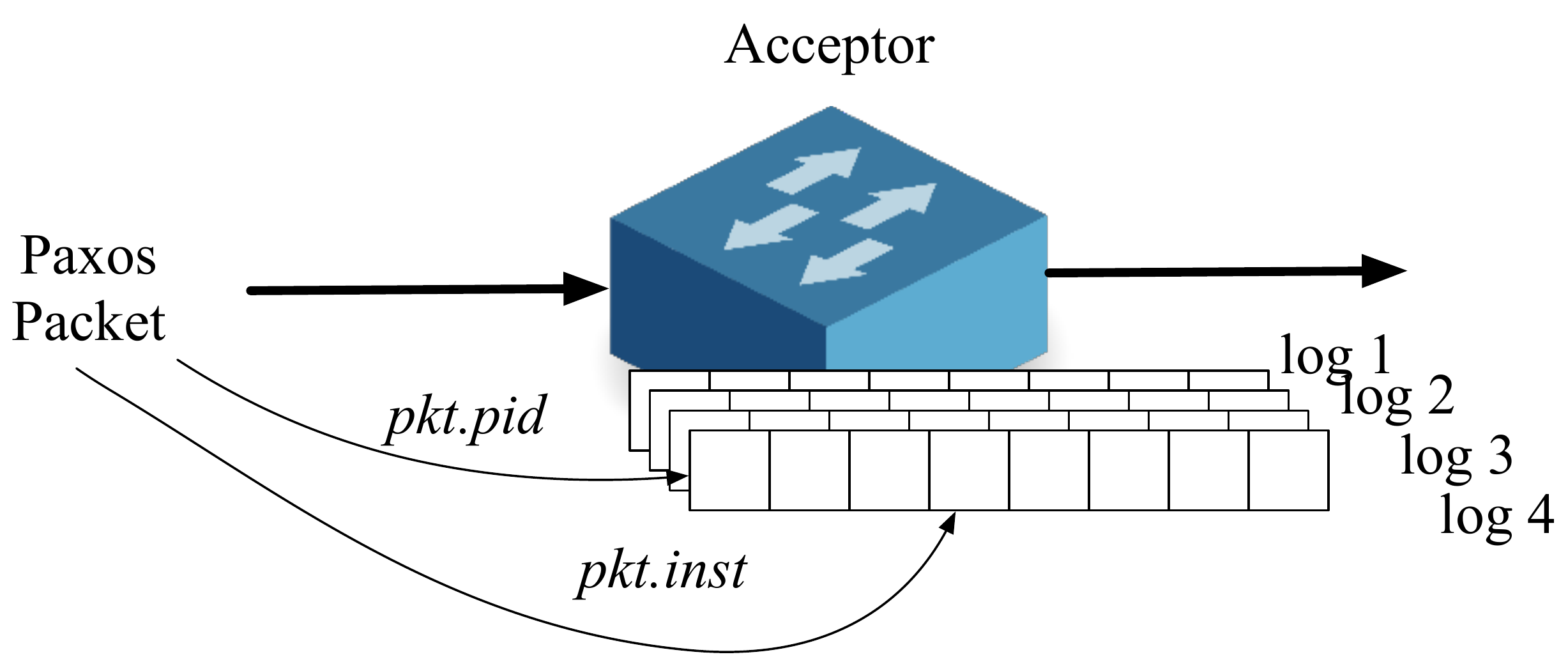}
\caption{Partitioned Acceptor log, indexed by partition id and instance number.}
\label{fig:partition}
 \vspace{-1em}
\end{figure}

Partitioned Paxos differs from traditional implementations
of Paxos in that it maintains multiple acceptor logs,
as illustrated in Figure~\ref{fig:partition}. Each log corresponds
to a separate partition, and each partition
corresponds to a separate shard of application state.
The acceptor log is implemented as a ring-buffer.

Algorithm~\ref{fig:acceptor} shows logic for an acceptor.  Acceptors
can receive either \texttt{PHASE1A} or \texttt{PHASE2A}
messages. Phase 1A messages are used during initialization, and Phase
2A messages trigger a vote. In both cases, the acceptor logic
must access the log for a particular partition and see if the
round number for the arriving packet is greater than the round
number stored at the switch. If not, the packet is dropped. Otherwise,
the switch modifying packet header fields and stored state,
depending on the message type.

\subsection{Resource Constraints}
\label{sec:resource-constraints}

Lamport's Paxos algorithm does not specify how to handle the ever-growing,
replicated log that is stored at acceptors. On any system, including
Partitioned Paxos, this can cause problems, as the log would require
unbounded disk space, and recovering replicas might need unbounded
recovery time to replay the log. To cope with log files, an
application using Partitioned Paxos must implement a mechanism to trim
the log~\cite{chandra07}.

In Partitioned Paxos, each acceptor maintains $P$ acceptor logs,
where $P$ is the number of partitions. Each log is implemented
as a ring buffer that can hold $I$ instance numbers. Thus,
the memory usage of Partitioned Paxos is $\mathcal{O}(P * I)$.
And, that the memory usage is inversely proportional to the
frequency of a log trim.

As will be described in
Section~\ref{sec:execution}, each partition of a replica must
track how many instance numbers have been agreed upon, and the largest
agreed upon instance number, $i$. When
the number of decided instances approaches $I$, the partition
must send a \texttt{TRIM} message to the acceptor. Upon
receipt of the \texttt{TRIM} message, the acceptor removes
 all state for instance numbers less than that $i$
Note that the \texttt{TRIM} message is sent as a data plane
command, not a control plane command.

With this static design, Partitioned Paxos must trim as frequently
as the slowest partition. An alternative design would allow for dynamic
partitioning, i.e., if all commands are for a single partition,
then that partition could use all available acceptor-dedicated switch memory.
However, this design requires scanning over the ring buffer to
clean the instance numbers for the trimmed partition.
This is difficult to implement in hardware.




\subsection{Failure Assumptions and Correctness}

Partitioned Paxos assumes that the failure of a leader or
acceptor does not prevent connectivity between the
consensus participants. As a result, it requires
that the network topology allows for redundant
routes between components, which is a common practice in data centers. In other respects,
the failure assumptions of Partitioned Paxos are the same
as in Lamport's Paxos. Below, we discuss how
Partitioned Paxos copes with the failure of a leader or
acceptor.


\paragraph{Leader failure.}

Paxos relies on a single operational leader to order messages.  Upon
the failure of the leader, proposers must submit proposals to a backup
leader.  The backup leader can be, for example, implemented in software.
If a proposer does not receive the response for a request after a
configurable delay, it re-submits the request, to account for lost
messages.  After a few unsuccessful retries, the proposer requests the
leader to be changed.

Routing to a leader or backup is handled in a similar fashion as the
way that load balancers, such as Maglev~\cite{eisenbud16} or Silk
Road~\cite{miao17-silkroad}, route to an elastic set of endpoints.  Partitioned
Paxos uses a reserved IP address to indicate a packet is
intended for a leader. Network switches maintain forwarding rules that
route the reserved IP address to the current leader.
Upon suspecting the failure of the hardware leader, a proposer submits
a request to the network controller to update the forwarding rules to
direct traffic to the backup.  A component that ``thinks'' it is the
leader can periodically check network controller that the reserved
leader IP address maps to its own address.  This mechanism handles
hardware leader failure and recovery.  To ensure progress, it relies
on the fact that failures and failure suspicions are rare events.

\paragraph{Acceptor failure.}

Acceptor failures do not represent a threat in Paxos, as long as a
majority of acceptors are operational.  Moreover, upon recovering from
a failure, an acceptor can promptly execute the protocol without
catching up with operational acceptors. Paxos, however, requires
acceptors not to forget about instances in which they participated
before the failure.

There are two possible approaches to meeting this requirement.
First, we could rely on always having a majority of operational acceptors
available. This is a slightly stronger assumption than traditional
Paxos deployments. Alternatively, we could require that acceptors
have access to persistent memory to record accepted instances.

Our prototype implementation uses the first approach, since the
network hardware we use only provides non-persistent SRAM.
However, providing persistent storage for network deployments
of Partitioned Paxos can be addressed in a number of ways. Prior work on
implementing consensus in FPGAs used on-chip RAM, and suggested that the
memory could be made persistent with a battery~\cite{istvan16}.
Alternatively, a switch could access non-volatile memory (e.g., an
SSD drive) directly via PCI-express~\cite{fpgadrive}.

%
%
%
%
%


\paragraph{Correctness}
Given this alternative interpretation of the Paxos algorithm,
it is natural to question if this is a faithful implementation
of the original protocol~\cite{lamport98}.
In this respect, we are aided by our P4 specification. In comparison
to HDL or general purpose programming languages, P4 is high-level and
declarative.  By design, P4 is not a Turing-complete language, as it
excludes looping constructs, which are undesirable in hardware
pipelines. Consequently, it is particularly amenable to verification
by bounded model checking.

We have mapped the P4 specification to Promela,
and verified the correctness using the SPIN model checker.
Specifically, we verify the safety property
of \emph{agreement}: the learners never decide on two separate
values for a single instance of consensus.

\section{Accelerating Execution}
\label{sec:execution}

\begin{figure}
\centering
 \includegraphics[width=0.7\columnwidth]{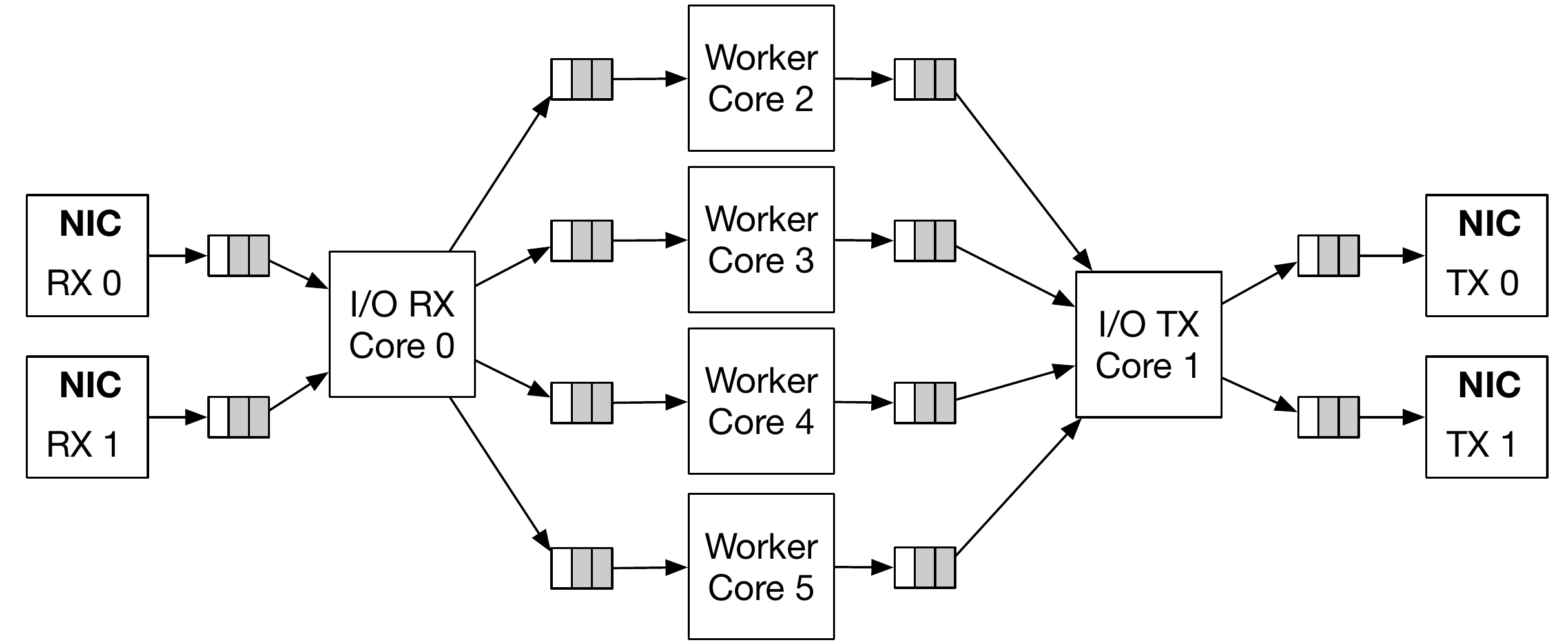}
\caption{Partitioned Paxos replica architecture.}
\label{fig:dpdk}
 \vspace{-1em}
\end{figure}

To accelerate the execution, Partitioned Paxos shards the application state at
replicas and assigns a worker thread to execute requests at each shard.
Our prototype currently only supports commands that access a single
shard. However, the approach can be generalized to support commands
that access multiple shards (i.e., multi-shard requests).

When a proposer submits a message with a request, it must include in
the message the shard (or shards) involved in the request (i.e., the partition id, \texttt{pid}).  Satisfying this constraint requires proposers to tell the read and write sets of
a request before the request is executed, as in, e.g., Eris~\cite{li2017eris}
and Calvin~\cite{thomson12}.  If this information is not available, a proposer can assume
a superset of the actual shards involved, in the worst case all shards.

Partitioned Paxos orders requests consistently across shards.
Intuitively, this means that if a multi-shard request $req_1$ is
ordered before another multi-shard request $req_2$ in a shard, then
$req_1$ is ordered before $req_2$ in every shard that involves both
requests.  Capturing Partitioned Paxos ordering property precisely is
slightly more complicated: Let $<$ be a relation on the set of requests such that
$req_1 < req_2$ iff $req_1$ is ordered before $req_2$ in some shard.
Partitioned Paxos ensures that relation $<$ is acyclic~\cite{coelho17}.

Every worker executes requests in the order assigned by Paxos.
Multi-shard requests require the involved workers to synchronize so that a single worker executes the request.
Therefore, multi-shard requests are received by workers in all involved shards.
Once a multi-shard request is received, the involved workers synchronize using a barrier and the worker with the lowest id executes the requests and then signals the other workers to continue their execution.
Supporting multi-shard commands introduces
overhead at the replica, which limits throughput.
Consequently sharding is most
effective when requests are single-shard and the load among shards is
balanced.

Note that each worker must track how many instance numbers have been
agreed upon, and the largest agreed upon instance number.  When the
number of agreed-upon instances exceeds a threshold, the worker must
send a \texttt{TRIM} message all acceptors. This message includes the largest agreed upon instance number and the partition
identifier. Upon receipt of this message, acceptors will trim their
logs for that partition up to the given instance number.


\subsection{Replica Architecture}

For a replica to realize the above design, there are two challenges
 that must be solved. First, the replicas must be able to process the
 high-volume of consensus messages received from the
 acceptors. Second, as the application involves writing to disk,
 file-system I/O becomes a bottleneck. Below, we describe how the
 Partitioned Paxos architecture, illustrated in Figure~\ref{fig:dpdk},
 addresses these two issues.

\paragraph{Packet I/O}
To optimize the interface between the network-accelerated
agreement and the application, Partitioned Paxos uses
a kernel-bypass library (i.e., DPDK~\cite{dpdk}), allowing
the replica to directly read packets from the server NIC.

Partitioned Paxos de-couples packet I/O from the application-specific
logic, dedicating a separate set of logical cores to each task.
The \emph{I/O Cores} are responsible for interacting with the NIC
ports, while the \emph{Worker Cores} perform the application-specific
processing.  The I/O Cores communicate with the Worker Cores via
single-producer/single-consumer lock-free queues (i.e., ring buffers).
This design has two key benefits.  First, the worker cores are
oblivious to the details of packet I/O activity. Second, the number of
cores dedicated to each task can be scaled independently, depending on
the workload and the characteristics of the replicated
application.

Figure~\ref{fig:dpdk} illustrates a deployment with one core dedicated to
receiving packets (I/O RX), one core dedicated to transmitting packets (I/O TX),
and four cores dedicated as workers. Both I/O cores
are connected to two NIC ports.

The I/O RX core continually poles its assigned NIC RX ring for arriving packets.
To further improve throughput, packet reads are batched. The
I/O RX core then distributes the received packets to the worker threads.
Our current implementation simply assigns requests using a static
partitioning (i.e., $\mbox{worker core} = \mbox{pkt.pid}\pmod{\mbox{NUM\_WORKER\_CORES}}$).
Although, more complex schemes are possible, taking into account the
workload. The only restriction is that all packets with the same \texttt{pid}
must be processed by the same worker.

Each Worker Core implements the Paxos replica logic---i.e., it receives a quorum
of messages from the acceptors, and delivers the value to the replicated application
via a callback. It is important to stress that this code is application-agnostic.
The application-facing interface would be the same to all applications, and
the same for any Paxos deployment.

\paragraph{Disk and File-System I/O}
The design described above allows Partitioned Paxos to process incoming packets
at a very high throughput. However, most replicated applications must also
write their data to some form of durable storage (e.g., HDD, SSD, Flash, etc.).
While different storage media will exhibit different performance characteristics,
our experience has been that the file system is the dominant bottleneck.

Unfortunately, many existing file system including ext4, XFS, btrfs,
F2FS, and tmpfs, have scalability bottlenecks for I/O-intensive
workloads, even when there is no application-level
contention~\cite{min16,balmau17}.  Therefore, to leverage the benefits of
sharding application state across multiple cores, Partitioned Paxos
uses a separate file-system partition for each application shard. In
this way, each file system has a separate IO scheduler thread.

\section{Evaluation}
\label{sec:evaluation}

\begin{figure}[t]
  \centering
  \includegraphics[width=0.4\columnwidth]{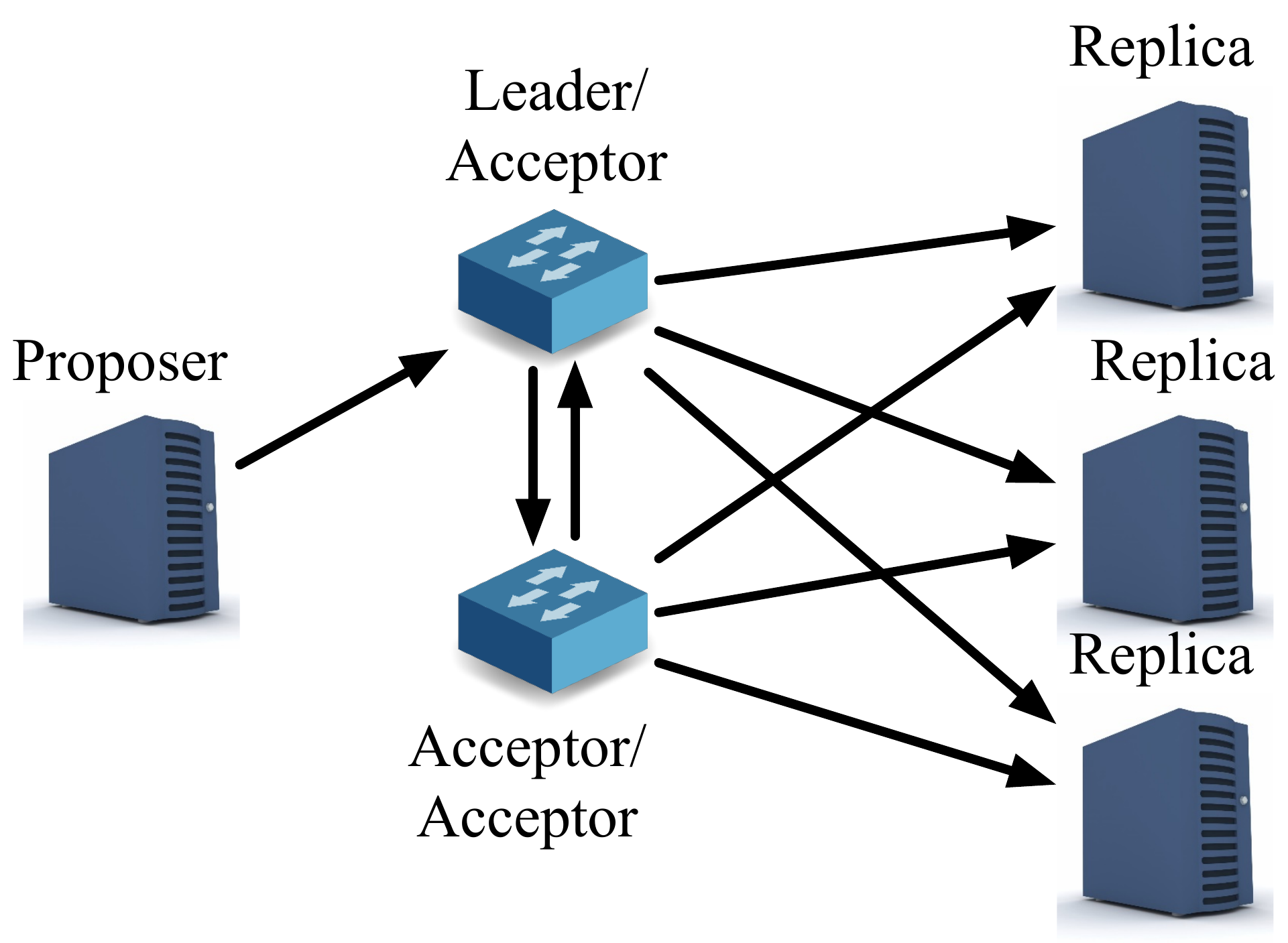}
  \caption{Topology used in experimental evaluation.}
  \label{fig:setup}
\end{figure}

Our evaluation of Partitioned Paxos explores four questions:
\begin{enumerate}

\item What is the absolute performance of individual Partitioned Paxos
components?
\item What is the resource overhead of in-network consensus on the network?
\item What is the end-to-end performance of Partitioned Paxos
as a system for providing consensus?
\item What is the performance under failure?
\end{enumerate}

As a baseline, we compare Partitioned Paxos with a software-based
 implementation, the open-source
\texttt{libpaxos} library~\cite{libpaxos}. Overall, the evaluation shows that Partitioned Paxos dramatically increases throughput and
reduces latency for end-to-end performance, when compared to
traditional software implementations.

\paragraph*{Implementation.}
\label{sec:implementation}
We have implemented a prototype of Partitioned Paxos. The switch code is written in P4~\cite{bosshart14},
and compiled to run on switches with Barefoot Network's Tofino ASIC~\cite{bosshart13}.
The replica code is written in C using the DPDK libraries.
We have an implementation of the Partitioned Paxos switch code that
targets FPGAs, as well. In the evaluation below, we focus
on the ASIC deployment.
All source code, other than the version that targets
Barefoot Network's Tofino chip, is publicly available with an
open-source license.

\paragraph{Experimental setup.}
In our evaluation, we used two different experimental setups.  Both
setups used 64-port, ToR switches with Barefoot Network's Tofino
ASIC~\cite{bosshart13}. The switches can be configured to run
at 10/25G or 40/100G.

In the first setup---used to test the absolute performance of
individual components---we used one switch configured to 40G per port. We followed a standard practice in industry
for benchmarking switch performance, a snake test. With a snake test,
each port is looped-back to the next port, so a packet passes through
every port before being sent out the last port. This is equivalent to
receiving 64 replicas of the same packet. To generate
traffic, we used a $2\times40Gb$ Ixia XGS12-H as packet sender and
receiver, connected to the switch with 40G QSFP+
direct-attached copper cables.
The use of
all ports as part of the experiments was validated, e.g., using
per-port counters. We similarly checked equal load across ports and
potential packet loss (which did not occur).


The second setup---used to test end-to-end performance and performance
after failure---was the testbed illustrated in Figure~\ref{fig:setup}.
Two Tofino switches were configured to run at 10G per port and logically
partitioned to run 4 Paxos roles.  One switch was a leader and an
acceptor. The second switch acted as two independent acceptors.

The testbed included four Supermicro 6018U-TRTP+ servers. One was used
as a client, and the other three were used as replicas.  The servers
have dual-socket Intel Xeon E5-2603 CPUs, with a total of 12 cores
running at 1.6GHz, 16GB of 1600MHz DDR4 memory and two Intel 82599 10
Gbps NICs. All connections used 10G SFP+ copper cables. The servers
were running Ubuntu 14.04 with Linux kernel version 3.19.0.

\subsection{Individual Components}

The first set of experiments evaluate the performance
 of individual Partitioned Paxos components deployed on a programmable ASIC.

\paragraph{Latency and throughput.}

We measured the throughput for all Paxos roles to be ~41 million 102
byte consensus msgs/sec per port.  In the Tofino architecture,
implementing pipelines of 16 ports
each~\cite{gurevich2016programmable}, a single instance of Partitioned
Paxos reached 656 million consensus messages per second.  We deployed
4 instances in parallel on a 64 port x 40GE switch, processing over
2.5 billion consensus msgs/sec. Moreover, our measurements indicate
that Partitioned Paxos should be able to scale up to 6.5 Tb/second of
consensus messages on a single switch, using 100GE ports.

We used Barefoot's compiler to report the precise theoretical latency
for the packet processing pipeline. The latency is less than
0.1~$\mu$s.  To be clear, this number does not include the SerDes,
MAC, or packet parsing components. Hence, the wire-to-wire latency
would be slightly higher.

Overall, these experiments show that moving Paxos into the forwarding
plane can substantially improve component performance


\begin{figure*}[t]
\centering
    \begin{subfigure}[t]{.475\textwidth}
        \centering
        \includegraphics[width=.9\columnwidth]{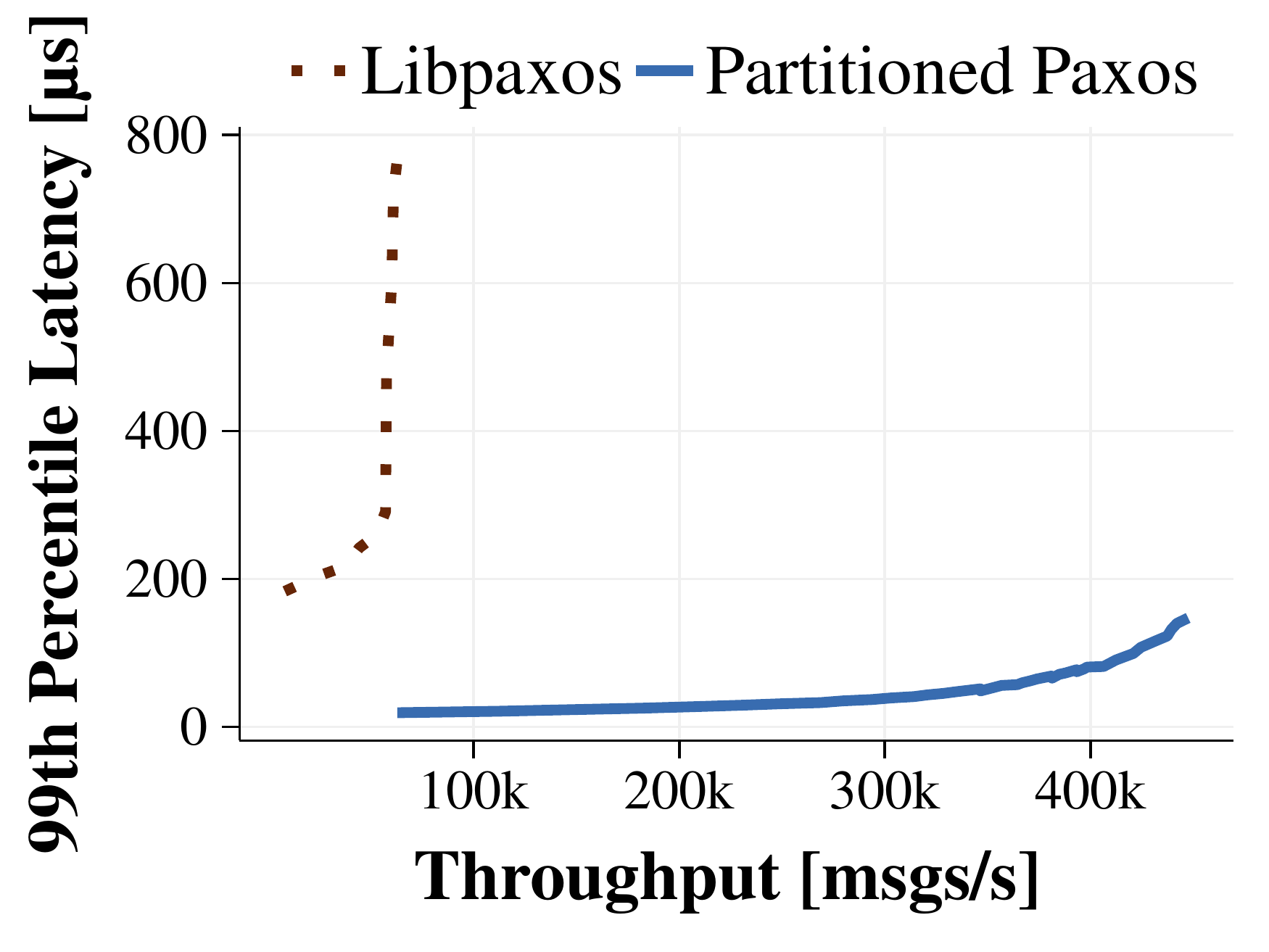}
        \caption{Tput vs. latency Noop.}
        \label{fig:tp_vs_latency_ack}
    \end{subfigure}%
    \hfill
    \begin{subfigure}[t]{.475\textwidth}
	   \centering
       \includegraphics[width=.9\columnwidth]{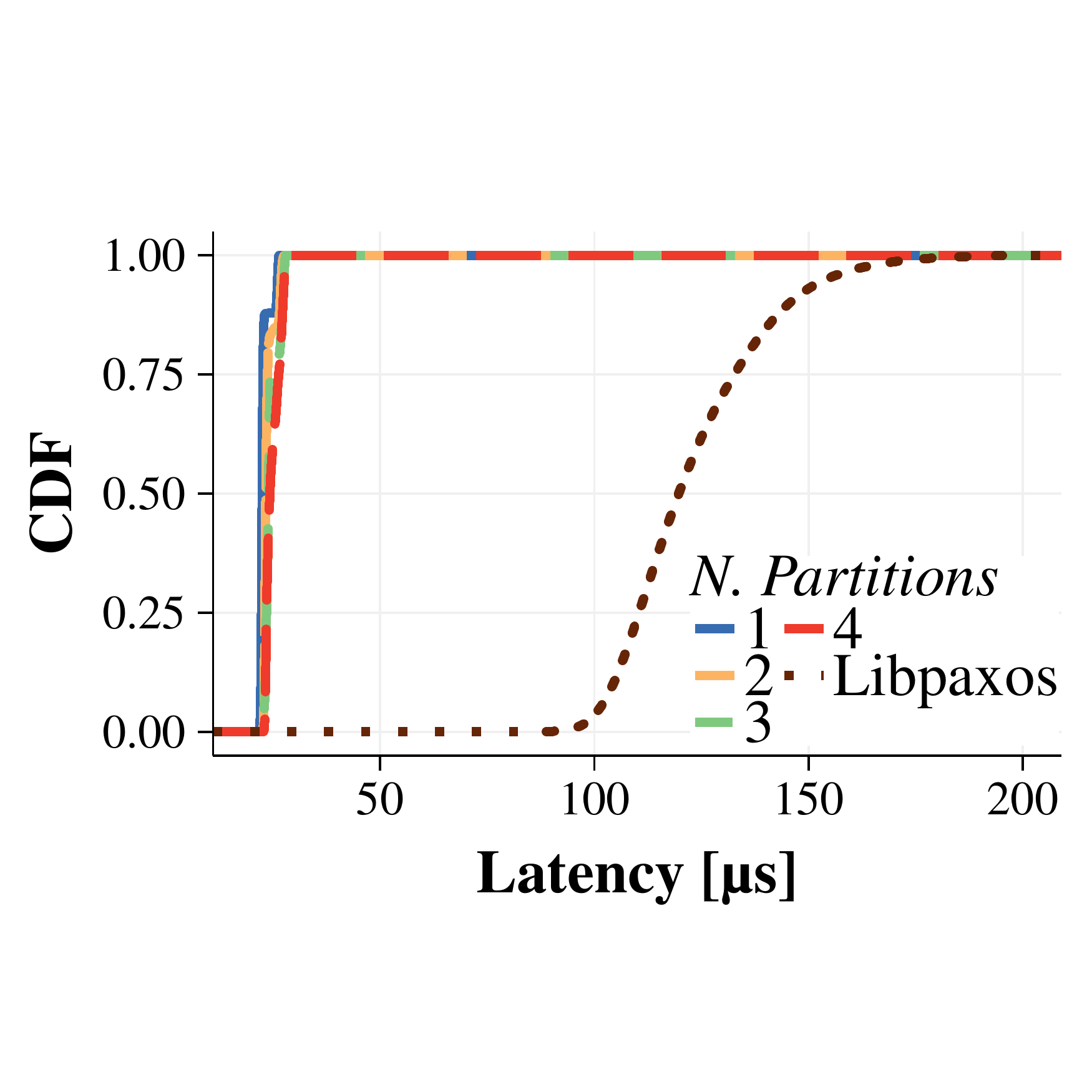}
       \caption{Latency CDF Noop.}
       \label{fig:ack-cdf}
    \end{subfigure}%
    \vskip\baselineskip
    \begin{subfigure}[t]{.475\textwidth}
        \centering
        \includegraphics[width=.9\columnwidth]{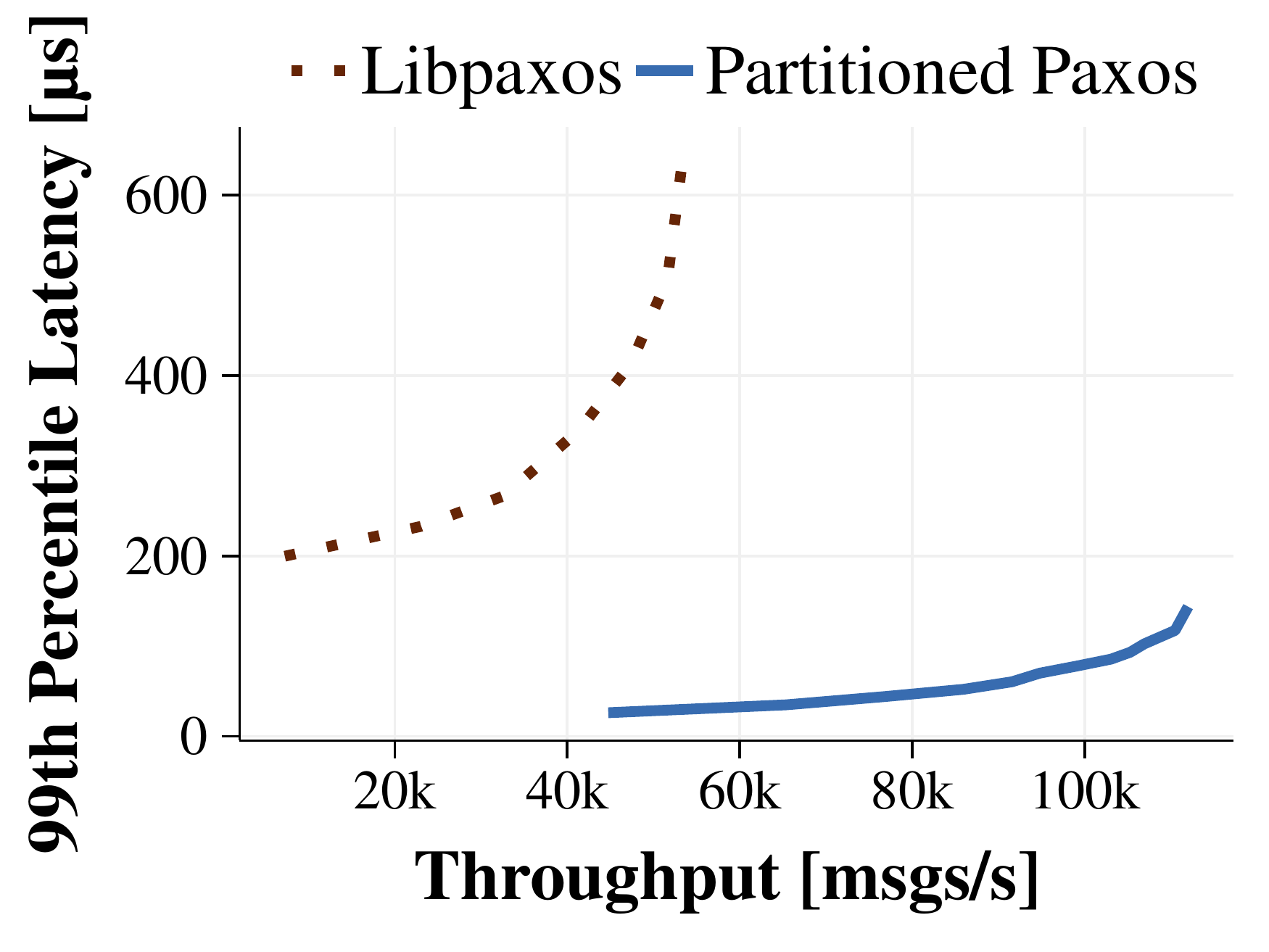}
        \caption{Tput vs. latency RocksDB.}
        \label{fig:tp_vs_latency_rocksdb}
    \end{subfigure}%
    \hfill
    \begin{subfigure}[t]{.475\textwidth}
    \centering
        \includegraphics[width=.9\columnwidth]{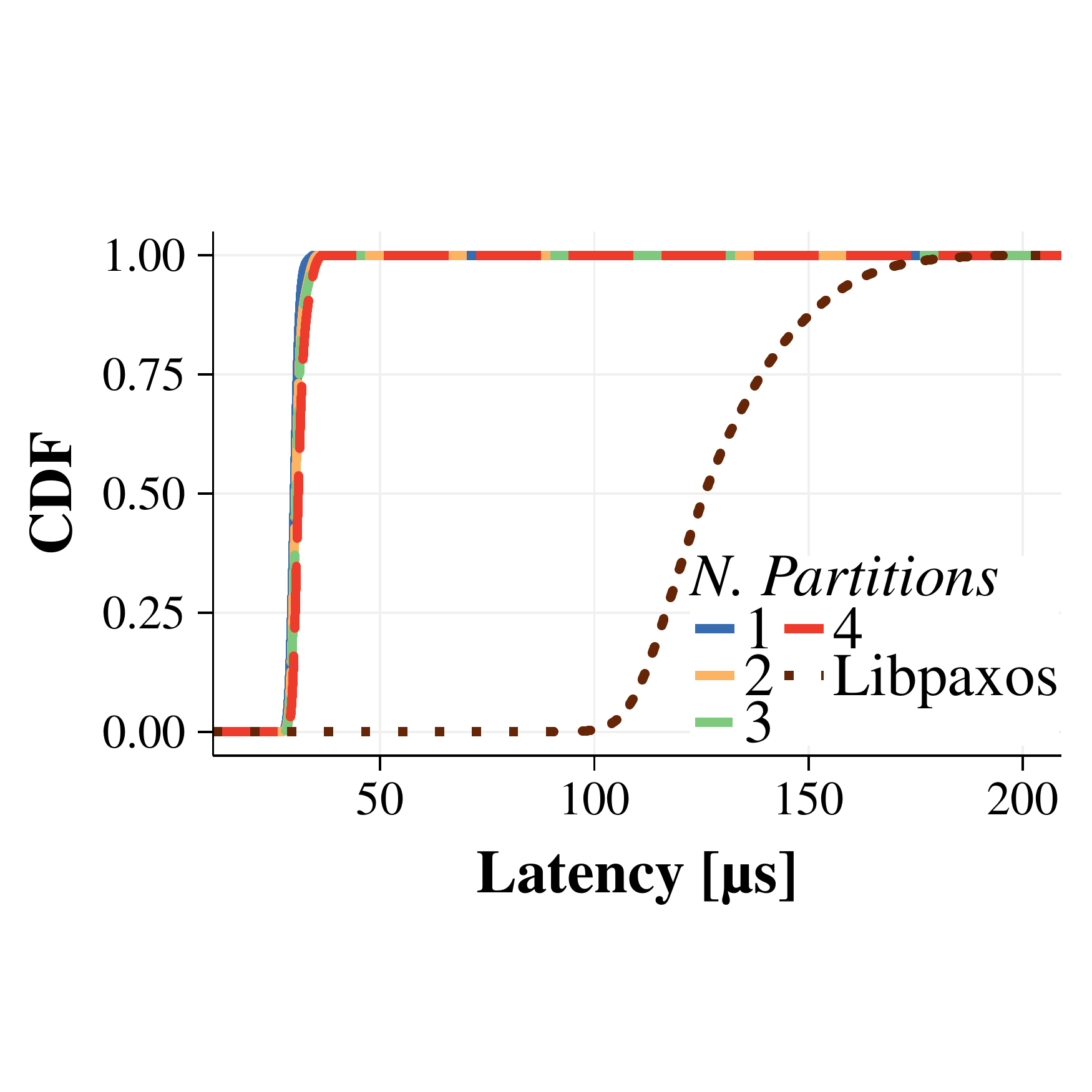}
        \caption{Latency CDF RocksDB.}
        \label{fig:rocksdb-cdf}
   \end{subfigure}%

\caption{The throughput vs. 99 percentile latency for a single partition of
the No-op application (a) and RocksDB (c). The latency at 50 percent of peak throughput for
No-op application (b) and RocksDB (d).}
\end{figure*}


\subsection{Resource Overhead}

The next set of experiments evaluates the resource overhead
and cost of running Partitioned Paxos on network
hardware.

\paragraph{Resources and coexisting with other traffic.}
We note that our implementation combines Partitioned Paxos logic with L2 forwarding.
The Partitioned Paxos pipeline uses less than 5\% of the available SRAM on Tofino, and no TCAM. Thus, adding Partitioned Paxos to an existing switch pipeline on a re-configurable ASIC would have a minimal effect on other switch functionality (e.g., storing forwarding rules in tables).

Moreover, the absolute performance experiment demonstrates how in-network computing can coexist with standard networking operation, without affecting standard network functionality, without cost overhead or additional hardware. Because the peak throughput is measured while the device runs traffic at full line rate of 6.5Tbps, there is a clear indication that the device can be used more efficiently, implementing consensus services parallel to network operations.

\paragraph{Power consumption}
A common criticism of in-network computing is that its power consumption overweights it performance benefits. To evaluate Partitioned Paxos power overhead, we compare the power consumption of Tofino running layer 2 forwarding, and layer 2 forwarding combined with Partitioned Paxos.
Due to the
large variance in power between different ASICs and ASIC
vendors~\cite{mellanox_spectrum}, we only report normalized power
consumption. Transceiver power consumption
is not accounted for (accounting for it would benefit Partitioned Paxos).
We again used a
``snake'' connectivity, which exercises all ports and enables testing Tofino at full capacity.

First, we measure the power consumption of both designs in idle, and find that it is the same, meaning that the Paxos program alone does not increase the power consumption. (i.e. activity is the one leading to additional power consumption). We then sent traffic at increasing rates.
The difference between the idle power consumption and maximum power consumption is only 2\%. While 2\% may sound like a significant number in a data center,
we note that the diagnostic program supplied with Tofino (diag.p4) takes 4.8\% more
power than the layer 2 forwarding program under full load.

\subsection{End-to-end Experiments}

Partitioned Paxos provides not only superior performance within the
 network, but also performance improvement on the application level,
 as we exemplify using two experiments.  In the first, the replicated
 application generates reply packets, without doing any computation or
 saving state.  This experiment evaluates the theoretical upper limit
 for end-to-end performance taking into account the network stack,
 but not other I/O (memory, storage) or the file system.  In the
 second experiment, we use Partition Paxos to replicate
 RocksDB~\cite{rocksdb}, a popular key-value store. RocksDB was
 configured with write-ahead logging (WAL) enabled.

As a baseline, both experiments compare Partitioned Paxos to
\texttt{libpaxos}. For the libpaxos deployment, the three replica servers
in Figure~\ref{fig:setup} also ran acceptor processes. One of the
servers ran a leader process.

\paragraph{No-op application}

In the first experiment, Server 1 runs a
multi-threaded client process written using the DPDK libraries.
Each client thread submits a message with the current timestamp
written in the value. When the value is delivered by the learner, a
server program retrieves the message via a deliver callback function,
and then returns the message back to the client.  When the client gets
a response, it immediately submits another message. The latency is
measured at the client as the round-trip time for each
message. Throughput is measured at the replica as the number of
deliver invocations over time.

To push the system towards higher a message throughput, we increased
the number of threads running in parallel at the client. The number
of threads, $N$, ranged from 1 to 12 by increments of 1. We stopped
measuring at 12 threads because the CPU utilization on the application
reached 100\%. For each value of $N$, the client sent a total of 10
million messages. We repeat this for three
runs, and report the $99^{th}$-ile latency and mean throughput.

Figure~\ref{fig:tp_vs_latency_ack} shows the throughput
vs. $99^{th}$-ile latency for Partitioned Paxos run on a
single partition. The deployment with libpaxos reaches
a maximum throughput of 63K. Partitioned Paxos
can achieve a significantly higher throughput at 447K, a $\times 7$
improvement. As well will see later, the throughput of Partitioned Paxos increases
even further as we add more partitions. Moreover, the latency
reduction is also notable.
 For Libpaxos, the latency at minimum
throughput is 183$\mu$s and at maximum throughput is 773$\mu$s.
The latency of Partition Paxos is only 19$\mu$s at 63K and 147$\mu$s at maximum throughput.

We measure the latency and predictability for Partitioned Paxos, and
show the latency distribution in Figure~\ref{fig:ack-cdf}.  Since
applications typically do not run at maximum throughput, we report the
results for when the application is sending traffic at a rate of 50\%
of the maximum. Note that this rate is different for libpaxos and
Partitioned Paxos: 32K vs. 230K, respectively.  Partitioned Paxos
shows lower latency and exhibits better predictability
than \texttt{libpaxos}: it's median latency is 22 $\mu$s, compared
with 120 $\mu$s, and the difference between 25\% and 75\% quantiles is
less than 1 $\mu$s, compared with 23 $\mu$s in \texttt{libpaxos}.
To add additional context, we performed the same experiment with
an increasing number of partitions, from 1 to 4.
We see that the latency for Partitioned Paxos has very little dependence on the
number of partitions.


\paragraph{RocksDB}

To evaluate how Partitioned Paxos can accelerate a real-world
database, we repeated the end-to-end experiment above for the no-op
experiment, but using
RocksDB instead as the application.  The RocksDB instances were
deployed on the three servers running the replicas. We followed the
same methodology as described above, but rather than sending dummy
values, we sent put requests to insert into the key-value store.  We
enabled write-ahead logging for RocksDB, so that the write operations
could be recovered in the event of a server failure.  It is important
to note that RocksDB was \emph{unmodified}, i.e., there were no
changes that we needed to make to the application.

Figure~\ref{fig:tp_vs_latency_rocksdb} shows the results. For
libpaxos, the maximum achievable throughput was 53K message / second.
For Partitioned Paxos---again, using a single partition---the maximum
throughput was 112K message / second. The latencies were also
significantly reduced. For libpaxos, the latency at minimum throughput
is 200$\mu$s and at maximum throughput is 645$\mu$s.  The latency of
Partitioned Paxos is only 26$\mu$s at 44K messages/second, and 143$\mu$s
at maximum throughput.

We measure the latency and predictability for Partitioned Paxos with replicated
RocksDB, and show the latency distribution in Figure~\ref{fig:rocksdb-cdf}.
As with the no-op server, we sent traffic at a
rate of 50\% of the maximum for each system. The rates were
23K for libpaxos and 65K for Partitioned Paxos. Again, we see that
Partitioned Paxos shows lower latency and exhibits better predictability
than \texttt{libpaxos}: it's median latency is 30 $\mu$s, compared
with 126 $\mu$s, and the difference between 25\% and 75\% quantiles is
less than 1 $\mu$s, compared with 23 $\mu$s in \texttt{libpaxos}.
As before, we repeated the experiment with 1, 2, 3, and 4 partitions.
The latency has very little dependence on the
number of partitions.



\paragraph{Increasing number of partitions}

Figures~\ref{fig:tp_vs_latency_ack} and~\ref{fig:tp_vs_latency_rocksdb} show the throughput
for Partitioned Paxos on a single partition. However, a key aspect of the design
of Partitioned Paxos is that one can scale the replica throughput by increasing the
number of partitions.

\begin{figure}[t]
\begin{minipage}[c]{0.45\linewidth}
\includegraphics[width=\linewidth]{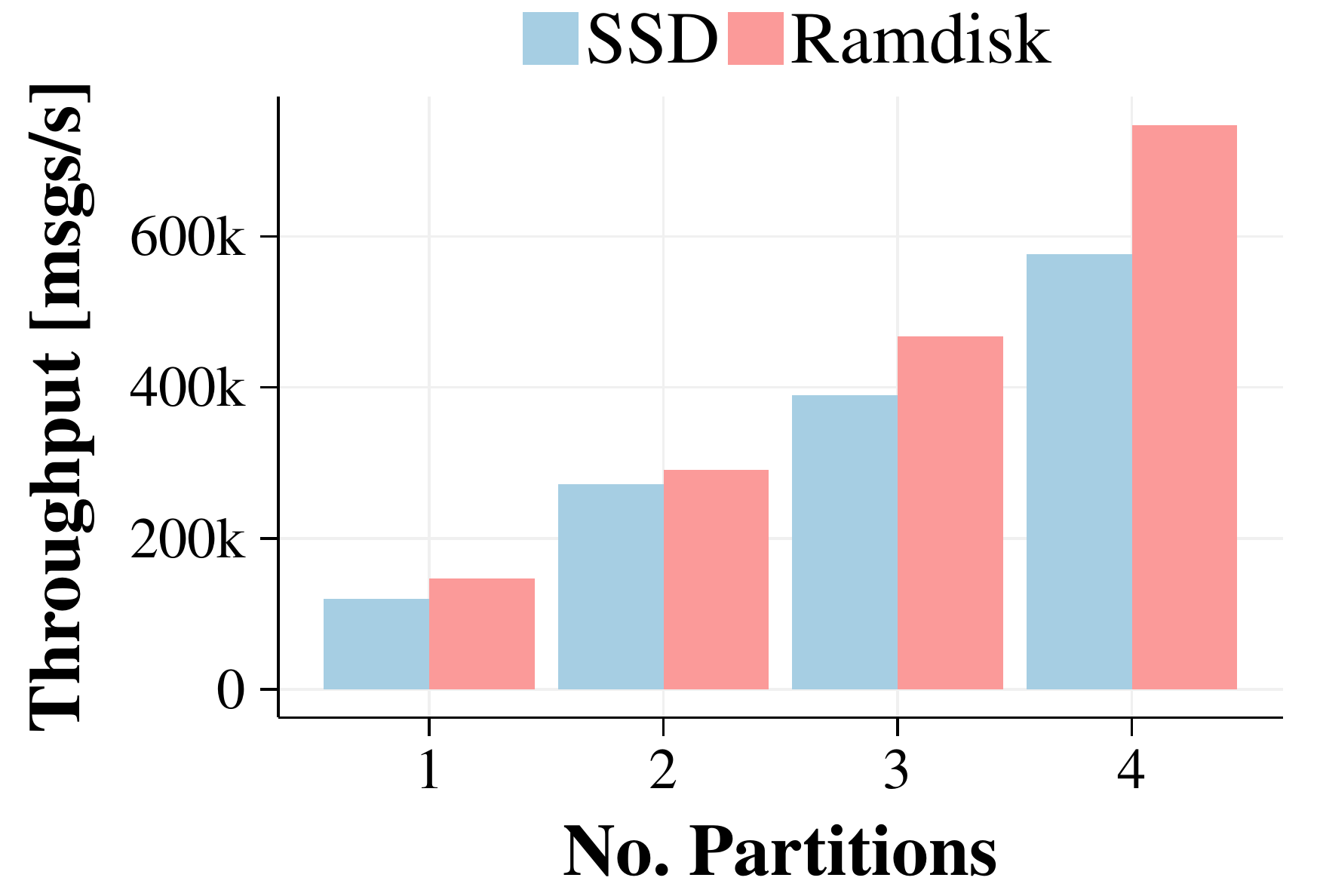}
\caption{Performance of Partitioned Paxos with RocksDB on Ramdisk and SSD.}
\label{fig:partitioning_rocksdb}
\end{minipage}
\hfill
\begin{minipage}[c]{0.45\linewidth}
\includegraphics[width=0.9\linewidth]{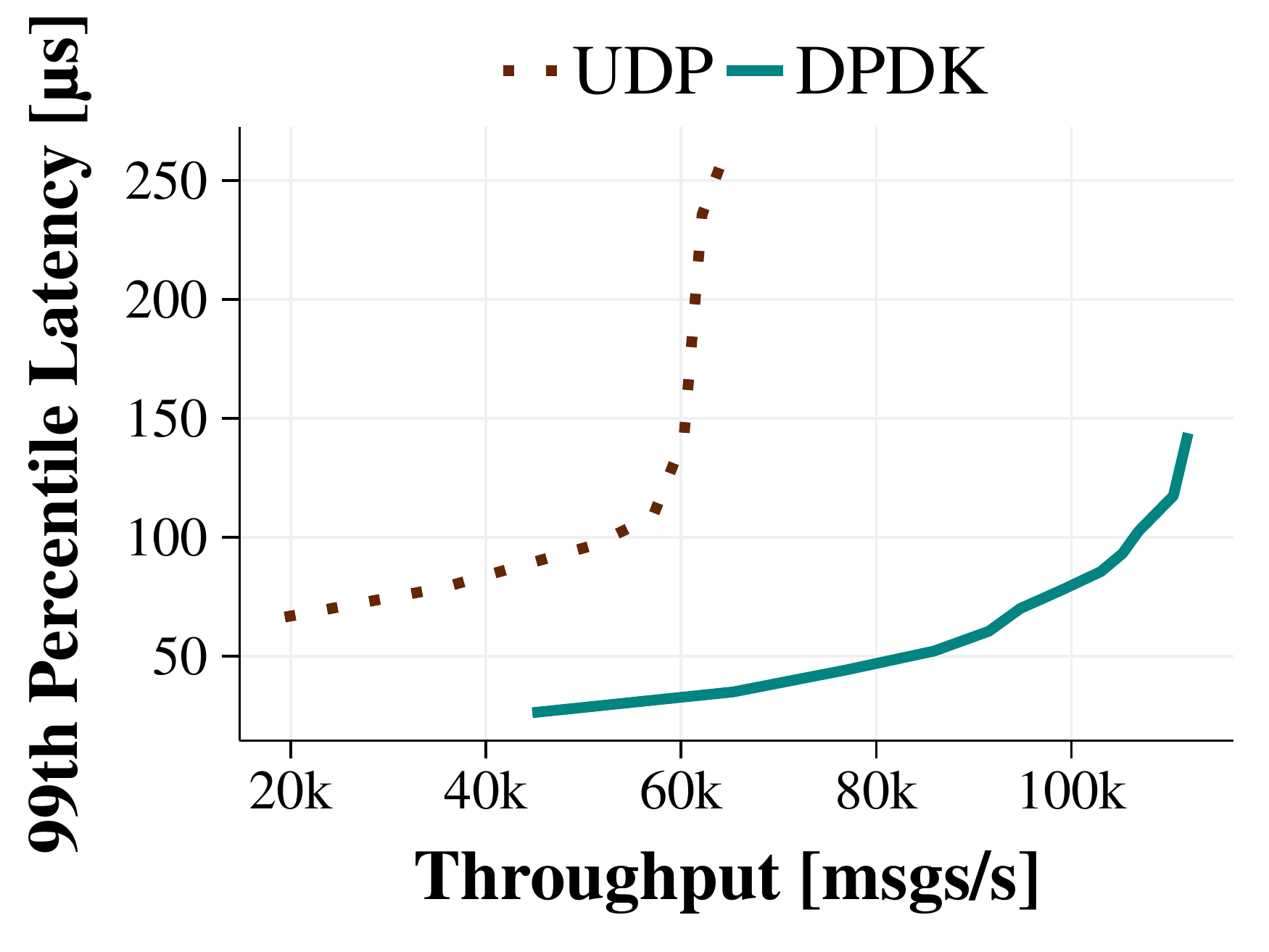}
\caption{Comparing throughput of replicated RocksDB using UDP socket vs. DPDK.}
\label{fig:udp_dpdk}
\end{minipage}
\end{figure}

Figure~\ref{fig:partitioning_rocksdb} shows the throughput of
 RocksDB with an increasing number of partitions, ranging
from 1 to 4. The figure shows results for different types of
storage media. For now, we focus on the results for SSD.

As we increase the number of partitions, the throughput increases linearly.
When running on 4 partitions, Partitioned Paxos reaches a throughput
of 576K messages / second, almost $\times$ 11 the maximum throughput
for libpaxos.



\paragraph{Storage medium}

To evaluate how the choice of storage medium impacts performance, we
repeated the above experiment using Ramdisk instead of an SSD.
Ramdisk uses system memory as a disk drive, i.e., it uses RAM instead
of SSD. As can be seen in Figure~\ref{fig:partitioning_rocksdb}, the
throughput increases linearly with the number of partitions. But, the
maximum throughput is much higher, reaching 747K messages / second.
This experiment eliminates the disk I/O bottleneck, and shows that
improving storage I/O can provide a 30\% performance improvement.  It
also shows that solving the storage bottleneck alone will not solve
all performance issues, i.e., it will not allow 1B packets at the
host.

\paragraph{DPDK}

To evaluate how much of the performance gains for Partitioned Paxos
can be attributed simply to the use of DPDK, we performed the
following experiment. We ran Partitioned Paxos on a single partition,
and replaced the DPDK library with a normal UDP socket. In both cases,
the replicas delivered requests to RocksDB for execution. The workload
consisted entirely of put requests.

Figure~\ref{fig:udp_dpdk} shows the results. We can see that DPDK
doubles the throughput and halves the latency.  For UDP, the latency
at minimum throughput (19K messages/second) is 66$\mu$s and at maximum
throughput (64K messages/second) is 261$\mu$s.  The latency of
DPDK is only 26$\mu$s at 44K messages/second and 143$\mu$s
at maximum throughput (112K messages/second).

\subsection{Failure Experiments}

\begin{figure}[t]
\centering
   \begin{subfigure}[t]{0.45\columnwidth}
     \centering
    \includegraphics[width=0.9\columnwidth]{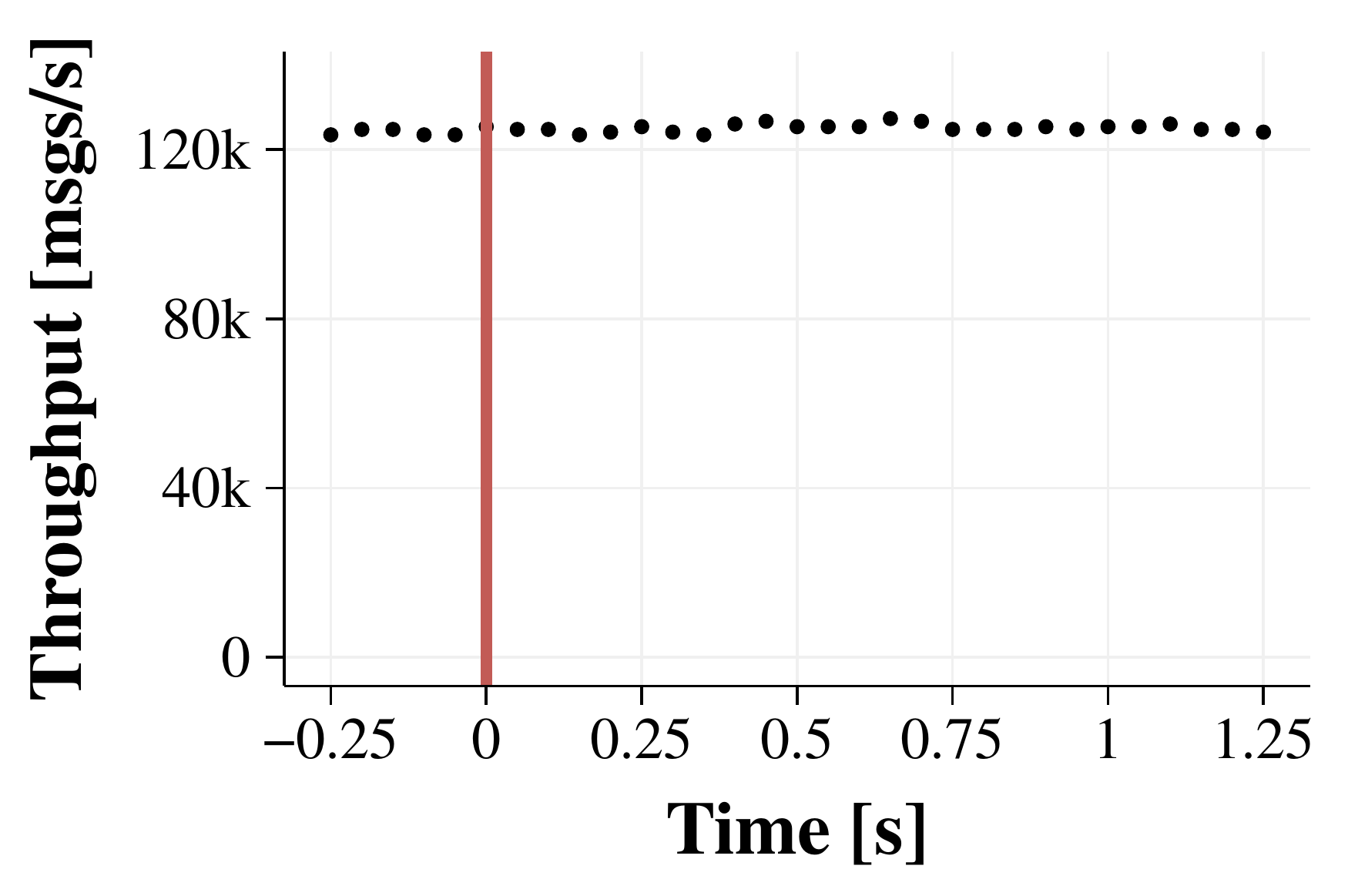}
    \caption{Acceptor failure}
    \label{fig:acceptor-failure}
   \end{subfigure}%
   \begin{subfigure}[t]{0.45\columnwidth}
     \centering
      \includegraphics[width=0.9\columnwidth]{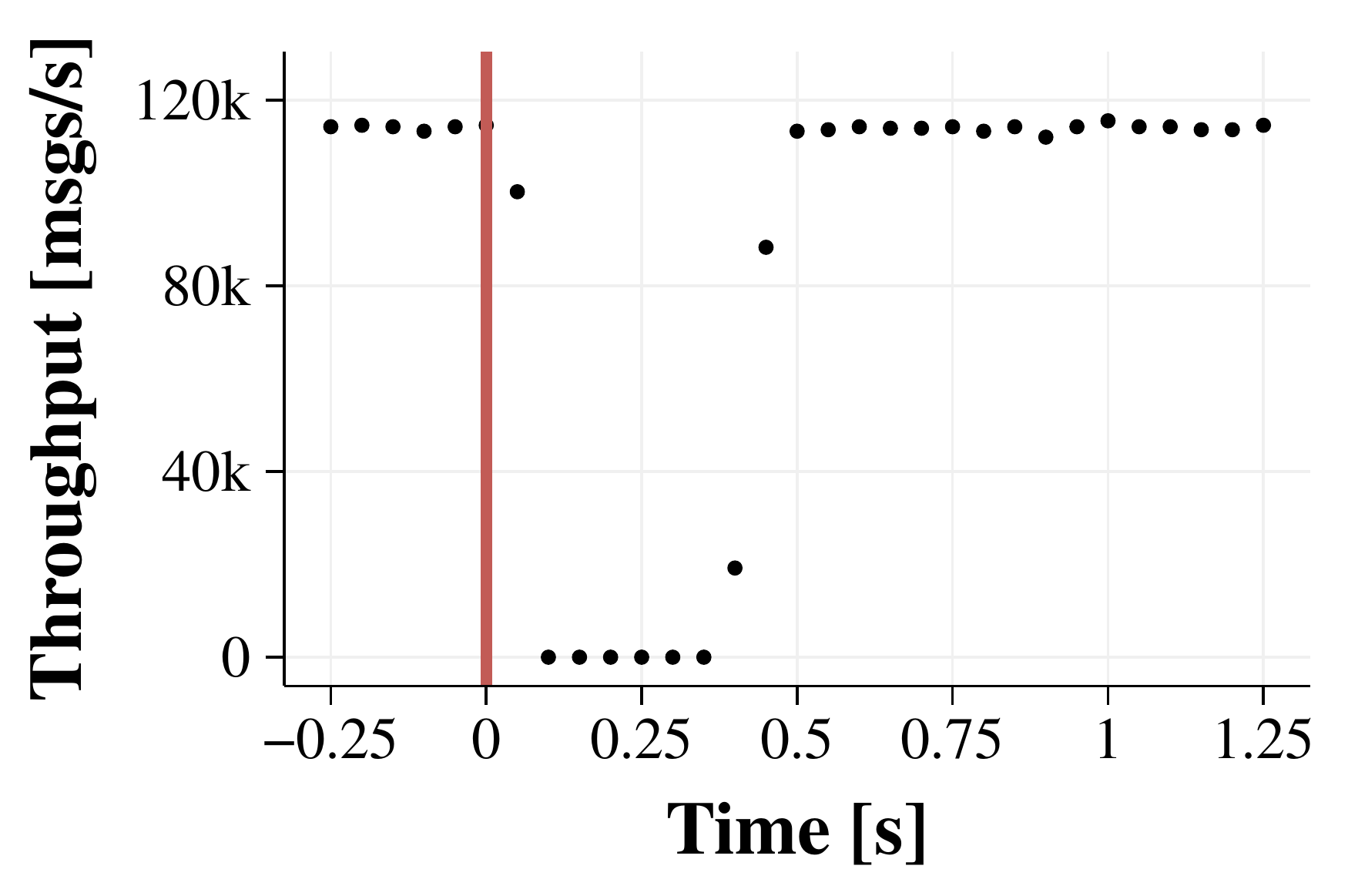}
        \caption{Leader failure}
        \label{fig:leader-failure}
   \end{subfigure}%
\caption{Throughput when (a) an acceptor fails, and (b) when FPGA leader is replaced by DPDK backup. The read line indicates the point of failure.}
\label{fig:failure}
 \vspace{-1em}
\end{figure}

To evaluate the performance of Partitioned Paxos after failures,
we repeated the latency and throughput
measurements under two different
scenarios. In the first, one of the three Partitioned Paxos acceptors fails. In
the second, the leader fails, and the leader is
 replaced with a backup running in software. In both the
graphs in Figure~\ref{fig:failure}, the vertical line indicates
the failure point. In both experiments, measurements were
taken every 50ms.

\paragraph{Acceptor failure}
To simulate the failure of an acceptor, we disabled the port between
the leader and one acceptor.
Partitioned Paxos continued to deliver messages at the same throughput,
as shown in Figure~\ref{fig:acceptor-failure}. In this single-partition
configuration, the bottleneck is the application.

\paragraph{Leader failure}
To simulate the failure of a leader, we disabled the leader logic on
the Tofino switch. After 3 consecutive retries, the proposer sends
traffic to a backup leader. In this experiment, the backup leader was
implemented in software using DPDK, and ran on one of the replicas.
The backup leader actively learns the chosen values from the primary
leader, so it knows the highest chosen Paxos instance.  The results,
appearing in Figure~\ref{fig:leader-failure}, show that the throughput
drops to 0 during the retry period. Again, because the application is
the bottleneck in the single-partition configuration. the system
returns to the peak throughput when the traffic is routed to the
backup leader. A DPDK-based implementation of a leader
can reach a throughput of $\sim$250K msgs/s.





\section{Related Work}
\label{sec:related}

Consensus is a well studied problem~\cite{lamport98,OL88,ongaro14,
CT96}. Many have proposed consensus optimizations,
including exploiting application semantics (e.g.,
EPaxos~\cite{moraru13}, Generalized Paxos~\cite{Lamport04}, Generic
Broadcast~\cite{PS99}), restricting the protocol (e.g., Zookeeper
atomic broadcast~\cite{reed08}), or careful engineering (e.g.,
Gaios~\cite{bolosky11}).


Recent work on optimizing consensus protocols rely on
two approaches: either they increase the strength of the assumptions that
a protocol makes about network behavior (e.g., reliable delivery,
ordered delivery, etc.). Or, they rely on increased support from
network hardware (e.g., quality-of-service queues,
support for adding sequence numbers, maintaining persistent state,
etc.).

Lamport's basic Paxos protocol
only assumes packet
delivery in point-to-point fashion and election of a non-faulty leader.
It also requires no modification to network forwarding devices.
Fast Paxos~\cite{lamport06} optimizes the protocol by optimistically
assuming a spontaneous message ordering~\cite{lamport06, PS02b,
PSUC02}. However, if that assumption is violated, Fast Paxos reverts to
the basic Paxos protocol.

NetPaxos~\cite{dang15} assumes ordered delivery, without enforcing the
assumption, which is likely unrealistic.  Speculative
Paxos~\cite{ports15} and NoPaxos~\cite{li16} use programmable
hardware to increase the likelihood of in-order delivery, and leverage
that assumption to optimize consensus
\`{a} la Fast Paxos~\cite{lamport06}.
In contrast, Partitioned Paxos makes few assumptions about the network behavior,
and uses the programmable data plane to provide high-performance.

Partitioned
Paxos differs from Paxos made Switch-y~\cite{dang16}
in several important ways. First, Partitioned Paxos
implements both Phase 1 and Phase 2 of the Paxos protocol in the
switch.  Second, it provides techniques for optimized Paxos replicas
(i.e., execution).  Third, Partitioned Paxos targets an ASIC
deployment, which imposes new constraints on the implementation.  And,
finally, Partitioned Paxos includes a quantitative evaluation of in-network consensus.

Istv\'{a}n et al.~\cite{istvan16} implement Zookeeper Atomic Broadcast (ZAB) in an FPGA.
ZAB uses TCP for reliable delivery. They also require that the
replicated application itself be implemented in the FPGA. In contrast,
Partitioned Paxos provides consensus for unmodified applications.

None of these prior systems address the problem of how
an application can take advantage of an accelerated consensus.
 Partitioned
Paxos builds on the idea of using the network data plane to
accelerate agreement, but also optimizes execution via state
partitioning and parallel execution.

In a separate, but related line of research,
Eris~\cite{li2017eris} and NOCC~\cite{jepsen18} use programmable
switches to accelerate transaction processing.


\section{Conclusion}
\label{sec:conclusion}

Partitioned Paxos significantly
improves the performance of agreement without additional hardware.
Moreover, it allows unmodified applications to leverage the
performance gains by sharding state and performing execution
in parallel. This is a first step towards a more holistic approach to designing
distributed systems, in which the network can accelerate services
traditionally running on the host.

\label{ConcPage}

\bibliographystyle{abbrv}
\bibliography{main}

\end{document}

%% file: abstract.tex
Consensus protocols are the foundation for building fault-tolerant,
distributed systems and services. They are also widely acknowledged as
performance bottlenecks. Several recent systems have proposed
accelerating these protocols using the network dataplane. But, while
network-accelerated consensus shows great promise, current systems
suffer from an important limitation: they assume that the network hardware
also accelerates the application itself. Consequently, they provide
a specialized replicated service, rather than providing a general-purpose
high-performance consensus that fits any off-the-shelf application.

To address this problem, this paper proposes Partitioned Paxos, a
novel approach to network-accelerated consensus. The key insight
behind Partitioned Paxos is to separate the two aspects of Paxos,
agreement and execution, and optimize them separately. First,
Partitioned Paxos uses the network forwarding plane to accelerate
agreement. Then, it uses state partitioning and parallelization to
accelerate execution at the replicas.
Our experiments show that 
using this combination of data
plane acceleration and parallelization, Partitioned Paxos
is able to provide at least $\times 3$ latency improvement and
 $\times 11$ throughput improvement for a replicated instance of a RocksDB key-value store.

%% file: algorithms/algorithm-leader.tex
\begin{algorithm}[t!]
  \small

\begin{distribalgo}[1]
  \INDENT{\textbf{Initialize State:}}
   \STATE instance[NumPartitions][1] $:=$ $\{0\}$
  \ENDINDENT
  \INDENT{\textbf{upon} receiving \mv{pkt(msgtype, inst, rnd, vrnd, swid, pid, value)}}

  \MATCH{\mv{pkt.msgtype}:}

  \CASE{\textsc{REQUEST}:}

  \STATE \mv{pkt.msgtype} $\leftarrow$ \textsc{Phase2A}
  \STATE \mv{pkt.rnd} $\leftarrow$ 0
  \STATE \mv{pkt.inst} $\leftarrow$ instance[pid][0]
  \STATE instance[pid][0] $:=$ instance[pid][0] + 1
  \MULTICAST {\mv{pkt}}
  \ENDCASE

  \DEFAULT{:}
  \DROP{\mv{pkt}}
  \ENDDEFAULT
  
  \ENDMATCH
\ENDINDENT

\caption{Leader logic.}
\label{fig:leader}
\end{distribalgo}

\end{algorithm}

%% file: algorithms/algorithm-acceptor.tex
\begin{algorithm}[t!]
\small

\begin{distribalgo}[1]

  \INDENT{\textbf{Initialize State:}}
   \STATE round[\textsc{NumPartitions}][\textsc{MaxInstances}] $:=$ $\{0\}$
   \STATE value[\textsc{NumPartitions}][\textsc{MaxInstances}] $:=$ $\{0\}$
   \STATE vround[\textsc{NumPartitions}][\textsc{MaxInstances}] $:=$ $\{0\}$
  \ENDINDENT

  \INDENT{\textbf{upon} receiving \mv{pkt(msgtype, inst, rnd, vrnd, swid, pid, value)}}
  \IF{\mv{pkt.rnd} $\geq$ round[\mv{pid}][\mv{pkt.inst}]}
  \MATCH{\mv{pkt.msgtype:}}

  \CASE{\textsc{Phase1A}:}
  \STATE round[\mv{pid}][\mv{pkt.inst}] $:=$ \mv{pkt.rnd}
  \STATE \mv{pkt.msgtype} $\leftarrow$ \textsc{Phase1B}
  \STATE \mv{pkt.vrnd} $\leftarrow$ vround[\mv{pid}][\mv{pkt.inst}]
  \STATE \mv{pkt.value} $\leftarrow$ value[\mv{pid}][\mv{pkt.inst}]
  \STATE \mv{pkt.swid} $\leftarrow$ swid
  \FORWARD{pkt}
  \ENDCASE

  \CASE{\textsc{Phase2A}:}
  \STATE round[\mv{pid}][\mv{pkt.inst}] $:=$ \mv{pkt.rnd}
  \STATE vround[\mv{pid}][\mv{pkt.inst}] $:=$ \mv{pkt.rnd}
  \STATE value[\mv{pid}][\mv{pkt.inst}] $:=$ \mv{pkt.value}
  \STATE \mv{pkt.msgtype} $\leftarrow$ \textsc{Phase2B}
  \STATE \mv{pkt.swid} $\leftarrow$ swid
  \FORWARD{\mv{pkt}}
  \ENDCASE

  \DEFAULT{:}
  \DROP{\mv{pkt}}
  \ENDDEFAULT

  \ENDMATCH
  \ELSE
  \DROP{\mv{pkt}}
  \ENDIF
\ENDINDENT

\caption{Acceptor logic.}
\label{fig:acceptor}
\end{distribalgo}
\end{algorithm}